\documentclass{jaa}
\usepackage[square]{natbib}
\usepackage[colorlinks]{hyperref}
\usepackage{xcolor, wasysym, ulem}
\usepackage[english]{babel}
\usepackage{epsfig}
\usepackage{epstopdf}
\usepackage{psfrag}
\usepackage{color}
\usepackage{graphicx}
\usepackage{fancyhdr}
\usepackage{graphics} 
\usepackage{amsmath}
\usepackage{xspace}
\usepackage{soul}
\usepackage{ulem}
\usepackage{lastpage}

\def\HI{{\rm HI}}

\def\V2{V_2}
\def\V2ij{V_{2ij}}

\def\V{\mathcal{V}}

\newcommand{\hi}{H{\sc i}\xspace}
\def\lsim{~\rlap{$<$}{\lower 1.0ex\hbox{$\sim$}}}

\def\gsim{~\rlap{$>$}{\lower 1.0ex\hbox{$\sim$}}}
\usepackage{soul,xcolor}

\begin{document}\sloppy

\title{Detecting galaxies in a large \hi spectral cube}

\author{Abinash Kumar Shaw\textsuperscript{1,2*}, Manoj Jagannath\textsuperscript{3}, Aishrila Mazumder\textsuperscript{4}, Arnab Chakraborty\textsuperscript{5}, \\Narendra Nath Patra\textsuperscript{4,6}, Rajesh Mondal\textsuperscript{7,8}, Samir Choudhuri\textsuperscript{9,10}}

\affilOne{\textsuperscript{1}Department of Physics, Indian Institute of Technology Kharagpur, Kharagpur  721302, India.\\}
\affilTwo{\textsuperscript{2}Astrophysics Research Centre, Open University of Israel, Ra'anana 4353701, Israel.\\}
\affilThree{\textsuperscript{3}Department of Electrical and Electronics Engineering, PES University, Bangalore 560085, India.\\}
\affilFour{\textsuperscript{4}Department of Astronomy, Astrophysics \& Space Engineering, Indian Institute of Technology Indore, Indore 453552, India.\\}
\affilFive{\textsuperscript{5}Department of Physics and McGill Space Institute, McGill University, Montreal, QC, Canada H3A 2T8.\\}
\affilSix{\textsuperscript{6}Astronomy and Astrophysics Division, Raman Research Institute, Sadashivanagar, Bengaluru - 560080, India.\\}
\affilSeven{\textsuperscript{7}Department of Astronomy and Oskar Klein Centre, AlbaNova, Stockholm University, Stockholm SE-10691, Sweden.\\}
\affilEight{\textsuperscript{8}Department of Astrophysics, School of Physics and Astronomy, Tel Aviv University, Tel Aviv 69978, Israel.\\}
\affilNine{\textsuperscript{9}School of Physics and Astronomy, Queen Mary University of London, London E1 4NS, U.K.\\}
\affilTen{\textsuperscript{10}Department of Physics, Indian Institute of Technology Madras, Chennai 600036, India.\\}
\date {}
\twocolumn[{\maketitle

\corres{\href{mailto:abinashkumarshaw@gmail.com}{abinashkumarshaw@gmail.com}}

\begin{abstract}
The upcoming Square Kilometer Array (SKA) is expected to produce humongous amount of data for undertaking \hi science. We have developed an MPI-based {\sc Python} pipeline to deal with the large data efficiently with the present computational resources. Our pipeline divides such large \hi 21-cm spectral cubes into several small cubelets, and then processes them in parallel using publicly available \hi source finder {\sc SoFiA-$2$}. The pipeline also takes care of sources at the boundaries of the cubelets and also filters out false and redundant detections. By comapring with the true source catalog, we find that the detection efficiency depends on the {\sc SoFiA-$2$} parameters such as the smoothing kernel size, linking length and threshold values. We find the optimal kernel size for all flux bins to be between $3$ to $5$ pixels and $7$ to $15$ pixels, respectively in the spatial and frequency directions. Comparing the recovered source parameters with the original values, we find that the output of {\sc SoFiA-$2$} is highly dependent on kernel sizes and a single choice of kernel is not sufficient for all types of \hi galaxies. We also propose use of alternative methods to {\sc SoFiA-$2$} which can be used in our pipeline to find sources more robustly.
\end{abstract} 

}]

\doinum{12.3456/s78910-011-012-3}
\artcitid{\#\#\#\#}
\volnum{000}
\year{0000}
\pgrange{1--\pageref{LastPage}}
\setcounter{page}{1}
\lp{\pageref{LastPage}}

\section{Introduction}

The baryonic matter content in the Universe is dominated by hydrogen. In its atomic form, neutral hydrogen (\hi) is one of the most reliable tracers of the formation and evolution of different structures in the Universe. The 21-cm line of neutral hydrogen can probe the evolutionary period of the Universe, starting from completely neutral at very high redshifts to ionized again over cosmic time. The ground state of neutral hydrogen atom consists of a proton and electron in the $1s$ orbital state. The spins of electron and proton are coupled and can be in a parallel or anti-parallel state. Hence, there are two spin states with an energy difference of about 5.87 $\mu$eV, where the higher spin state corresponds to the parallel spins of proton and electron. The hyperfine transition of the atom between these two spin states can be mediated by the absorption or emission of a photon of wavelength $\sim$ 21 cm (1420 MHz). However, the spontaneous emission coefficient for this hyperfine states is very small, $A_{10}=2.85 \times 10^{-15}$s$^{-1}$. Due to this, it is challenging to detect the \hi\ line emission from individual galaxies beyond redshift $z\sim 0.4$,  with modern telescopes. But the large fraction of baryons ($\sim 75\%$) is neutral hydrogen, and due to which the 21 cm line signal of neutral hydrogen becomes an excellent tracer of the structure formation. This redshifted \hi\ 21 cm line can be used to map the Universe in 3D.

Galaxy evolution studies in the last decades have majorly focused on understanding different physical mechanisms which drive the growth of galaxies, starting with cold atomic hydrogen (\hi\ gas) through the cold interstellar
medium (ISM; e.g., where the bulk of dense molecular hydrogen, $\mathrm{H_{2}}$, resides) to star formation \citep{Walter2020ApJ}. 
Galaxy evolution includes baryonic processes like gas infalling onto galaxies to form neutral atomic hydrogen (\hi\ gas), which is then converted to the molecular state ($\mathrm{H_{2}}$), and, finally, the conversion of $\mathrm{H_{2}}$ to stars \citep{Chowdhury2020}. Thus, understanding the evolution of neutral atomic and molecular hydrogen and that of stars is required to understand galaxy evolution. Previous studies have shown that the comoving star formation rate density (SFRD) rises towards the end of the epoch of reionization ($z \lesssim 6$) and peaks in the redshift range $z \sim 1-3$ and then declines by an order of magnitude from $z \sim 1$ to today \citep{Walter2020ApJ}. However, the neutral \hi\ mass density ($\Omega_{\mathrm{H{\sc I}}}$) does not show any significant evolution over cosmic time.  \citet{Chowdhury2020} shows that accretion of  gas onto galaxies at $z \leq 1$ may have been insufficient to sustain high star-formation
rates in star-forming galaxies and likely to be the cause of the decline in the cosmic star-formation rate density at redshifts below one. However, the evolution of cold neutral gas of large samples of galaxies during $z \sim 0-3$ is still needed to be constrained with more sensitive observation to understand the global flow of gas onto galaxies and probe the history of star formation in the Universe. In addition, the \hi\ 21 cm emission line is being used to measure the rotation curve of the galaxies, which is one of the direct probes of the dark matter \citep{Rubin1970}.

There has been significant development in the theoretical understanding of the role and evolution of \hi in the Universe using semi-analytical modeling and hydrodynamic simulations (the readers are referred to Section 2.2 of \citet{Blyth:2015UX} for a summary of recent advancements). From the observation side as well, there have been surveys like HIPASS \citep{hipass2001},  ALFALFA \citep{alfafa2005} that have provided valuable information on global \hi properties and galaxies in general. Such surveys have also provided the best estimates of \hi mass function (HIMF). However, sensitivity limitations have confined surveys to the local Universe (z$\sim$0.3). There have been successful surveys of the 21-cm signal in the local Universe. These have produced \hi masses of many galaxies \citep{zwaan2005,jaffe2013,catinella2014, fernandez2016, jones2018}. There have been some studies beyond the low-redshift regime, where either intensity mapping (for example \citet{Chang2010}, \citet{Masui_2013}) or averaging over known galaxy spectra (\hi stacking, e.g., \citet{jayaram2001, lah2007, Kanekar_2016, rhee2018, Bera_2019, Chowdhury2020}) has been done. The CHILES survey with the VLA is another example of \hi survey in the local Universe \cite{Chiles2013}. This survey reported the highest redshift detection of  \hi 21 cm emission signal, so far,  from a galaxy at $z\sim0.376$ in the COSMOS field \citep{fernandez2016}.

The Square Kilometer Array (SKA) and its precursor facilities like MeerKAT and ASKAP promise to address the gaps left by current telescopes with their enhanced sensitivities and collecting areas. For instance, the MHONGOOSE survey using the MeerKAT \citep{mhongoose} is a deep survey using 30 local disk and dwarf galaxies to study \hi distribution. This study is expected to address several scientific questions like the effect of cold accretion, the relation between gas and star formation, dynamics of magnetic fields, etc. On the other hand, the LADUMA survey with the MeerKAT is targeting the observation of the Chandra Deep Field South (CDFS) to use \hi for studying the evolution of galaxy over $\sim$70\% of the age of Universe \citep{LADUMA}. The survey forecasts predict the detection of sources up to $z\leq1.45$ \cite{LADUMA}. The major LADUMA science questions are: (i) dependence of HIMF on environment density and redshift (ii) obtain direct measurements of cosmic \hi density at high redshifts, (iii) study dependence of galactic \hi masses on the stellar mass and host dark matter halo mass as well as other properties as a function of surroundings and redshift (iv) time evolution of baryonic Tully-Fisher relations, etc. The survey is ongoing, and the first data release is expected in 2022. MeerKAT is also conducting the \hi emission project for the MIGHTEE survey \citep{mightee_hi}. It will be a deep blind survey with medium width to study neutral hydrogen emissions up to $z\sim0.6$. The survey aims to trace how the neutral gas in the galaxies evolved over the past 5 billion years. The survey volume is designed to cover massive groups and galaxy clusters while observing low-mass galaxies ($<10^{8}\,{\rm M}_\odot$) as well. Several surveys-- WALLABY, FLASH, DINGO are also being undertaken by the ASKAP telescope towards \hi science. The WALLABY survey \citep{wallaby} is targeting to detect around half a million galaxies in the local Universe $z\leq0.26$, with the primary science goals including obtaining a census of gas-rich galaxies in the Local Group, studying the impact of the environment on galaxy evolution, and refinement of cosmological parameters. The FLASH \citep{flash} survey aims to probe higher redshifts ($0.4<z<1.0$) to explore the \hi content in the Universe. FLASH will search for absorption lines against the continuum of bright radio sources to produce a database of \hi-absorption selected galaxies rich in cool, star-forming gas. It also aims to provide insights into kinematic information for models of gas accretion and jet-driven feedback in radio-loud active galactic nuclei. DINGO survey aims to provide deep legacy data sets for \hi emission out to $z\sim0.4$ \citep{dingo}. The fundamental science objectives for this survey are tracing the evolution of \hi mass density, studying the cosmic web, and understanding how \hi evolves along with other constituents in galaxies over the past 4 Gyr.

The \hi surveys with different SKA precursor facilities will create an excellent inventory of the role of \hi in various stages of the formation and evolution of galaxies. However, SKA will provide detailed information to characterize the physical processes governing galaxy evolution. SKA will also offer higher angular resolution and sensitivities than currently possible \citep{Blyth:2015UX}. SKA will address the issues with the measurement of the HIMF. The HIMF is well measured for the local Universe, and the planned deep surveys with the precursors will enable its measurement up to $z<0.6$. SKA will push this measured redshift to $z\sim1$ to provide more stringent constraints on galaxy evolution. It will also enable probing the cosmic neutral gas density to higher redshifts than currently possible. These measurements will also be useful to study the relationship between \hi mass and stellar mass, star formation rate (SFR), morphology, etc., down to lower \hi masses. Gas inflow from the IGM and its removal from the galaxy plays a crucial role in the formation and evolution of galaxies. Observations for detecting the processes involved in the gas inflow are currently sensitivity limited. SKA will increase the sensitivity by almost two orders of magnitude \citep{Blyth:2015UX} with high spatial resolution enabling the detection of gas infalling into the galaxy. With the SKA, it will also be possible to study tidal interactions and associated events in a large sample of galaxies - their prevalence, environment and time dependence, etc. The Tully-Fisher relation (TFr) \citep{Tully1977} describes the correlation between a galaxy’s intrinsic luminosity and the distance-independent width of its global \hi profile. It is one of the most valuable tools to obtain distances to galaxies in the local Universe, where peculiar velocities dominate over the Hubble flow. At the same time, its statistical properties are useful for constraining simulations probing galaxy formation and evolution. The SKA precursors will improve observations of the TFr by the improved spatial resolution of the \hi kinematics, thus improving TF-based distances in the local Universe. While the SKA itself is unlikely to improve upon this significantly, its sensitivity and angular resolution will make it possible to use TFr for probing galaxy evolution with cosmic time. This will be achieved by using its slope, scatter, and zero-point as a function of cosmic environment and time (redshift) and other properties such as morphological type and SFR. 

Thus, with its unprecedented sensitivity, SKA will be the most powerful instrument to study neutral hydrogen in galaxies. However, one limitation is that there are still no clear indications on how far in redshift beyond $z\sim1$ SKA can probe reliably. Additionally, any SKA survey will come with the inherent problem of data analysis. The high data volume anticipated with SKA observations will make analysis challenging. Powerful supercomputing facilities would be required to process the typical raw data, and these will not be accessible to the community in general. Even after some reduction stages, the data will typically be $\sim$1 TB in size for \hi galaxy science. Thus, there is a requirement to develop efficient algorithms that will be able to deal with such high volumes of data using reasonable computing power. Towards this end, the SKA usages are presented regularly with Science Data Challenges to familiarize the scientific community with the expectations from this telescope. Recently, the SKA-DC2\footnote{\url{https://sdc2.astronomers.skatelescope.org/}}, where the participants were presented with 1 TB data cube across a $20$ square degree sky area to find and characterize the \hi content of galaxies. 

The challenge saw participation from several teams across the globe and provided some truly novel methods to deal with the \hi galaxy science problem. There were several groups from India, and our group, Team Spardha, was able to complete the full challenge. This paper summarizes our efforts and describes the methodology adopted for data analysis. We have used a combination of traditional methods and modern techniques like machine learning. Within the time limit of the challenge, we were unable to explore all our options completely. Thus we are currently testing new methods, with the ultimate goal of providing an efficient pipeline for performing \hi galaxy science. The paper is arranged as follows: In sec. \ref{sims}, we mention the details of the simulated data cube,  sec. \ref{method} describes the method we have developed so far to detect galaxies from an image cube, in sec. \ref{results} we show the results, and finally, we present a summary and conclusions in sec. \ref{summary}. 
 
 
\section{Simulations}
\label{sims}

To characterize our algorithm and routines, we use simulated data provided by the SKA Office (SKAO). This data set was made publicly available as part of the SKA-Data Challenge 2 (SKA-DC2). In particular, we used their development data set, which has a volume of $40$ GB.
Note that our analysis routines are generic and implemented using MPI-based parallel algorithm, which can easily be extended to larger data volumes ($\sim$ TB). Here in this section, we briefly describe how these \hi data cubes are generated. For more details, we refer the readers to \citet{skadc2}.

The simulated \hi cubes represent mock data equivalent to SKA observations. In this case, the spectral cubes contain \hi signal from galaxies within observing frequency range between $950$ MHz to $1150$ MHz. This corresponds to a redshift interval of $z = 0.235 - 0.495$. A channel width of $30$ kHz (velocity width of $\sim 6$ km/s) is used, producing $\sim 6700$ spectral channels across the cube. The development data cube covers a field of view $\sim 1$ square degrees with a beam size of $7$ arcsec and a pixel size of $2.8$ arcsec. This corresponds to a total comoving volume $V=645833.2\,{\rm Mpc}^3$ in the sky. The data cube was generated to simulate $2000$ hours of SKA observation. Accordingly, the cube has noise, systematic and other imperfections (e.g., Radio Frequency Interference (RFI), residual noise, deconvolution error, etc.) added. 

The \hi signal in these cubes originates from the distribution of galaxies in the universe. Hence, to simulate them, a proper prescription is used to mimic their 3D distribution, physical properties, and other observational effects. A detailed description of the same can be found in \citet{skadc2} and \citet{bonaldi21}. Here, we briefly describe the steps adopted to produce the \hi cubes. 

The \hi spectral cubes are simulated in three broad steps. First, considering the distribution of galaxies in the universe, a source catalog is generated. Next, a sky model is produced incorporating model observations from nearby galaxies at the source positions on the catalog. Finally, this sky model is conditioned with telescope response and artifacts to create the final cube.  

The catalog of the \hi sources is generated by sampling the \hi mass function. This provides the average number of galaxies per unit volume within some \hi mass range. The \hi mass function is adopted from the ALFALFA survey \citep{jones18} and can be given by  

\begin{equation}
    \phi ({\rm M}_{\rm HI},z) = \ln(10) \thinspace \phi_* \thinspace \left( \frac{{\rm M}_{\rm HI}}{{\rm M}_*(z)} \right)^{\alpha + 1} \exp{\left({-\frac{{\rm M}_{\rm HI}}{{\rm M}_*(z)}}\right)},
\end{equation}

\noindent where, $\phi ({\rm M}_{\rm HI}, z)$ represents the number of galaxies per ${\rm Mpc}^3$ having \hi mass of ${\rm M}_{\rm HI}$ at a redshift $z$. ${\rm M}_*$ is the knee mass above which the number of galaxies decreases sharply. $\alpha$ is the power-law index, and $\phi_*$ is the normalization constant. A redshift-dependent knee mass is used to include mild evolution of the \hi mass function as a function of redshift, 

\begin{equation}
    \log {\rm M}_*(z) = \log {\rm M}_* + 0.075 \times z.
\end{equation}

Several physical properties (e.g., \hi diameter, total integrated flux, velocity widths, etc.) of the galaxies are generated using the \hi mass from the catalog and different scaling relations observed in the local universe \citep[see, e.g.,][]{duffy12,wang16}. Note that the source catalog used to generate the $40$ GB data cube contains a total of $11091$ \hi galaxies. These galaxies contain \hi masses varying within the range $1.4\times 10^8\,{\rm M}_{\odot}$ to $6.6\times 10^{10}\,{\rm M}_{\odot}$, and their velocity width ranges between $30.29\, {\rm km/s} - 891.87 \,{\rm km/s}$.

Along with the \hi, the SKA observations will also contain signals from continuum sources. These continuum sources could originate in the \hi bearing galaxy itself (due to star formation) or through other galactic processes (e.g., AGN activity). However, their presence in the data cube is crucial and should be appropriately modeled. A continuum catalog of extragalactic sources was generated using Tiered Continuum Radio Extragalactic Continuum Simulation \citep[T-RECS, ][]{bonaldi19}. A population in this continuum catalog would be associated with the \hi sources (originated by SFR). Hence, the continuum catalog is cross-matched against the \hi catalog to identify sources compatible with both the \hi and continuum properties in the catalogs. Due to the inclusion of full diversity, this cross-identification would leave many sources in both the catalogs, which will not have any match. This results in three categories of sources, `continuum only' (sources do not have an \hi counterpart, e.g., AGNs, or \hi is too faint to be detected), `regular' (sources having both \hi and continuum detected), and `\hi only' (sources do not have a continuum, e.g., low SFR galaxies). Further, to mimic the real distribution, these sources are clustered using the dark matter distribution from P-Millennium simulation \citep{baugh19} as a backbone.

Model intensity distribution must be built based on the source properties at individual source locations. Thus, actual observations are used from WSRT Hydrogen Accretion in LOcal GalaxieS (HALOGAS) survey \citep{fraternali02,oosterloo07,heald11} and the THINGS survey \citep{walter08} to prepare an atlas of templates. For every galaxy in the catalog, a galaxy in the atlas is identified whose physical properties match the best. The template was then appropriately smoothed and sampled to maintain the consistency of the data cube at the galaxy location (e.g., pixel size, linear resolution).  

After the model observations are placed at the galaxy locations in the cube, telescope response should be incorporated to simulate actual observation. The telescope beam is computed using the SKA-MID configuration. The $uv$ distribution is produced using one minute of sampling over $8$ hours of observation. Ideally, this $uv$ distribution should be used to produce dirty beams for each channel. However, some of the $uv$ data could be flagged due to artifacts. The noise in the data mainly has two components, thermal noise, and artifacts due to Radio Frequency Interference (RFI). The thermal noise is calculated considering an equivalent observing time of $2000$ hours. For RFI contribution, first, a baseline power level (noise floor) is obtained using observation of the MeerKAT telescope. This noise floor is then scaled up by a user-defined value contributing to RFI and distributed amongst all the baselines. To mimic real-life data analysis, a fraction of this RFI is then flagged. This leaves a non-uniform $uv$ distribution at the end, used to estimate the final dirty beam. This dirty beam is used to introduce several imaging-related artifacts, e.g., residual sky image after deconvolution, etc. \citet[see, ][for more details]{skadc2}. The final spectral cube was made by adding the actual \hi signals and different artifacts originated due to imperfect observations and imaging.

Using the prescription mentioned above, SKA-DC2 team simulated the $40$ GB \hi spectral cubes which is used in characterizing our \hi detection techniques. The following section describes the method we use to identify galaxies in an \hi spectral cube.


\section{Methods}\label{method}

We used the publicly available 3D \hi\  Source-Finding Application ({\sc SoFiA}) on data cubes to identify galaxies and extract their properties \citep{Serra_2015}. Although this software has been developed to detect \hi\ galaxies in  Widefield ASKAP L-band Legacy All-sky Blind Survey (WALLABY),  {\sc SoFiA} works on any data cube independent of the telescope and observed spectral line \citep{Serra_2015}. However, the requirement of a  large memory footprint and comparatively slow speed of {\sc SoFiA} makes it challenging to work on large data volumes. To overcome these issues with large data volume, a recently updated version is released for  {\sc SoFiA}-$2$ \citep{Westmeier_2021}. It re-implements most of the algorithms of {\sc SoFiA} in the C-programming language and uses OpenMP for multithreading of the most time-critical algorithms and parallelization to significantly speed up the processing of large data volumes. {\sc SoFiA}-$2$ can process a single 800 GB WALLABY data cube in minutes on a modest number of computing nodes \citep{Westmeier_2021}. We used this latest version {\sc SoFiA}-$2$ in our work. Here we shall briefly describe the algorithms of {\sc SoFiA}-$2$, for a detailed description, please see \citet{Serra_2015, Westmeier_2021} and references therein.  

\begin{figure*}
    \centering
    \includegraphics[width=0.8\textwidth]{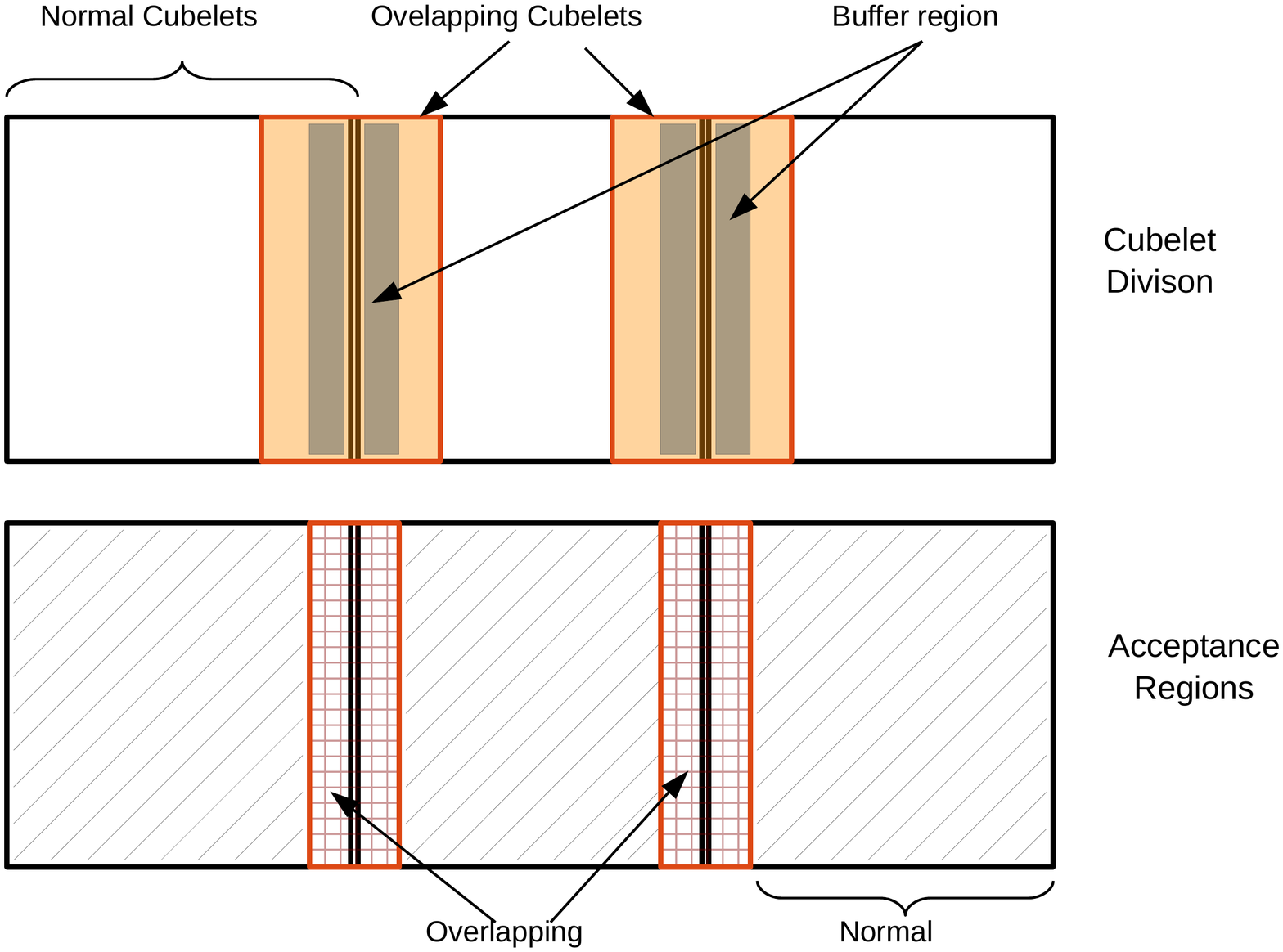}
    \caption{Shows the 2D projection of the schematic division of the data into the {\it Normal} and {\it Overlapping} cubelets along one of the axis (top row) and the corresponding Acceptance regions (bottom row).}
    \label{fig:spardha}
\end{figure*}

{\sc SoFiA}-$2$ works on the stokes-I data cube in Flexible Image Transport System (FITS) format \citep{Pence_2010}. It is possible to process different sub-regions of the entire cube simultaneously. {\sc SoFiA}-$2$ has an auto-flagging algorithm implemented which can flag spatial pixels and spectral channels for which the noise deviates from the median noise across all pixels or channels by a user-specified multiple of the RMS as estimated from the more robust median absolute deviation. This helps in identifying corrupted data due to radio frequency interference (RFI) and flag them. Note that, {\sc SoFiA}-$2$ works only on constant noise across the data cube to apply threshold algorithm to detect the sources. Hence, if the noise level is not constant across the data cube, {\sc SoFiA}-$2$ first normalizes the noise and then applies the source-detection algorithm.  

There are two source finding algorithms implemented inside {\sc SoFiA}-$2$: a simple
threshold finder and the smooth and clip ({\bf S+C}) finder. The threshold algorithm applies an absolute or relative (with respect to the noise) flux-based cut-off to identify sources. This has been rarely used unless the data is preconditioned. The most powerful and widely used algorithm is the {\bf S+C} finder. This algorithm iteratively smooths the data cube on multiple spatial and spectral scales to obtain significant emission above a user-specified detection threshold on each scale \citep{Serra_2012}. {\sc SoFiA}-$2$ estimates both positive and negative flux values above the threshold. The negative flux values are used to estimate the false detection rate and provide the statistical reliability of the catalog with positive flux values \citep{Westmeier_2021}. 

After detecting the pixels above the threshold, the software uses a simple friends-of-friend algorithm to link all pixels within a merging radius specified by the user. It assigns a unique identifier to each detection. The linking and merging of pixels happen in 3D, \textit{i.e.}, in spatial and frequency directions. A minimum and maximum filters can also be applied to filter out the noise artifacts, otherwise considered actual sources. {\sc SoFiA}-$2$ determine the reliability of each detection and use this
information to automatically remove detections from the source catalogue that are deemed unreliable \citep{Dickinson_2004, Westmeier_2021}. {\sc SoFiA}-$2$ compares the number density of detections with negative and positive total flux in a 3D parameter space made up of the peak
flux density, the summed flux density, and the mean flux density across the source and then estimates the reliability of the source based on the user-specified reliability threshold. This helps to reduce the false detections in the source catalog. In addition to the reliability threshold, the user can also set a signal-to-noise threshold. All detections with an integrated signal-to-noise ratio (SNR) below that threshold will be discarded as unreliable irrespective of their measured reliability. {\sc SoFiA}-$2$ also determines the basic source properties, like peak flux, integrated flux density, position angle of the galaxy, line width, etc., and write those in the output catalog \citep{Serra_2015, Westmeier_2021}. The strategy we employed to detect and extract the source properties efficiently from a data cube with {\sc SoFiA}-$2$ is discussed below.

We have developed a {\sc Python} based pipeline which starts with dividing the development data cube of size $40$ GB  into several small cubelets. We analyze all the cubelets in parallel using an MPI-based implementation, where we have run parallel instances of {\sc SoFiA}-$2$ \citep{Westmeier_2021} on each cubelet to find the sources. We have tuned the parameters of {\sc SoFiA}-$2$ to maximize the number of detected sources. A total of $118$ cubelets were analyzed, which can be categorized into two groups, namely, (1) Normal cubelet and (2) Overlapping cubelet. We divide the whole data cube into consecutive blocks of equal dimensions, which we indicate as Normal cubelets as shown by black outlined boxes in Figure \ref{fig:spardha}.

We separately define and analyze the Overlapping cubelets, which encompass regions in the adjacent normal cubelets and are centered at their common boundary as shown by orange boxes in Figure \ref{fig:spardha}. This is done for detecting the sources which fall at the common boundaries. The normal cubelets and the overlapping cubelets always have buffer zones (e.g., blue regions for Normal cubelets) around their faces to avoid confusion between the sources detected near their common boundaries. We conservatively set the width of buffer zones based on the physically motivated values of the spatial (on the sky plane) and frequency extent of typical galaxies scaled at the desired redshifts. We choose the maximum extent of the galaxy on the sky plane to be $\sim 80~{\rm kpc}$ which roughly corresponds to $10$ pixels on the nearest frequency channel. Therefore the buffer region was set to be twice \textit{i.e.} $20$ pixels. Hence we make the Overlapping region $4\times 20=80$ pixels wide. Similarly along the frequency direction, galaxies can have extent $\approx 72$ channels which corresponds to a line-width $\sim 500~{\rm km/s}$. Therefore the widths of the buffer region and overlapping region along the frequency-axis are $144$ and $288$ channels, respectively. We always accept any source whose center is detected within the cubelet but not in the buffer zone, as demonstrated in the bottom row of Figure \ref{fig:spardha}. The acceptance regions of the cubelets (normal and overlapping) are defined in a particular way so that they span the whole data cube contiguously when arranged accordingly. Although this increases the computation a bit due to analyzing some part of data multiple times (once in normal cubelet and once in associated overlapping cubelet), it ensures no common source is present in the list after this step. Analyzing cubelets is the most time-consuming part of our pipeline. We analyze $118$ cubelets on $32$ cores in parallel in around $20$ minutes.

Next, we use the physical equations to convert the {\sc SoFiA}-$2$ catalogue into the physical units and discard the bad detections, \textit{i.e.} sources having \textit{NaN} values in the columns or with negative flux values, etc. In the final stage we put a cap on $w_{20}$ (line-width) to discard the detections with unusual velocity-width. Motivated by the physical models/observations of the galaxies, we have conservatively accepted the sources having $w_{20}\in [60,\, 500]~{\rm km/s}$. We finally arranged the catalogue in the descending order of the flux values. Based on our experience with the `Development Data Cube', for which the exact source properties are known, we chose around the top 35\% of the sources to evaluate our pipeline's efficiency.  

\section{Results}\label{results}

Using the method described above, we try to detect \hi sources in the spectral cube. For the current study, we use {\sc SoFiA}-$2$ as our primary tool for detecting the galaxies in the data cube.

\begin{figure*}
     \centering
     \includegraphics[scale=0.2]{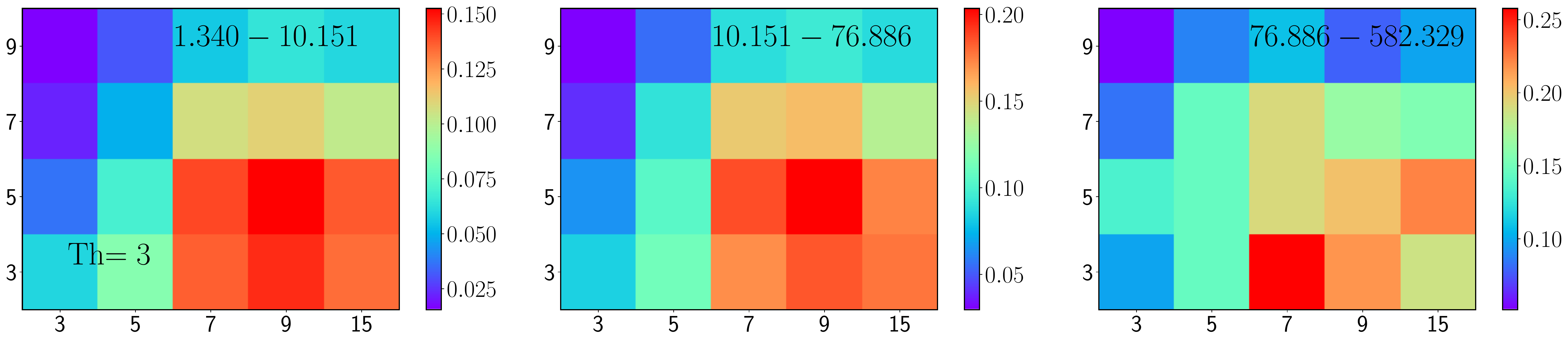}
     \hspace*{-0.55cm}
     \includegraphics[scale=0.2]{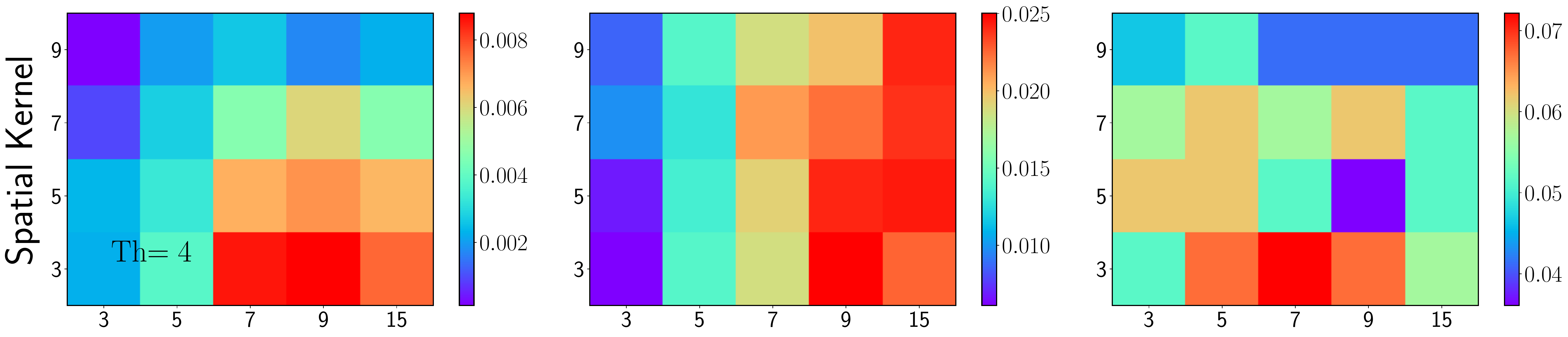}
     \includegraphics[scale=0.2]{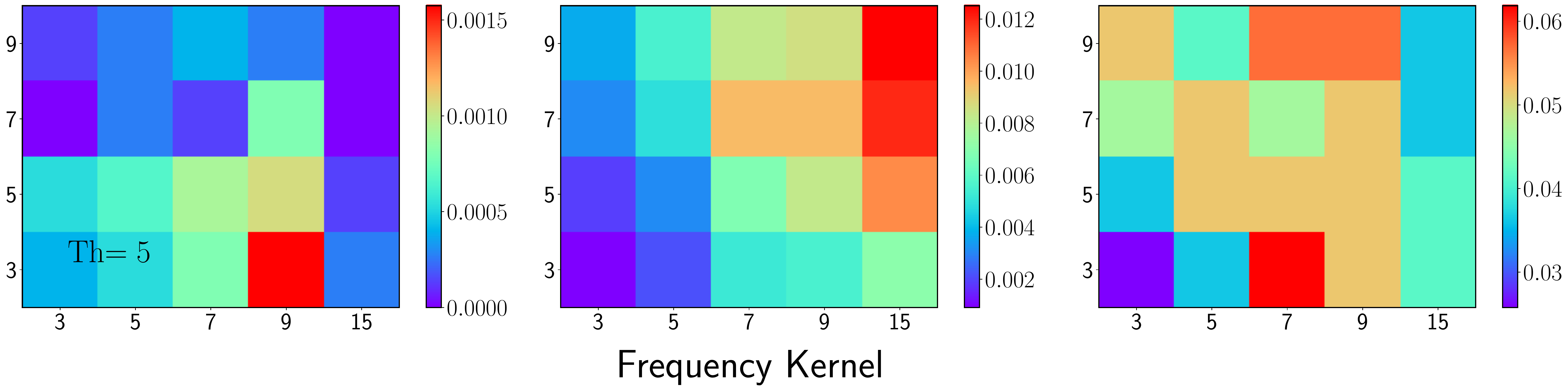}
     \caption{Shows the fraction of detected sources as a function of Spatial kernel size and Frequency kernel size. The three columns represent the three integrated flux bins with the ranges shown in the top-right corner in the top row.  The three rows corresponds to the three threshold values as stated in the first column panel.}
     \label{fig:eff}
 \end{figure*}

Though {\sc SoFiA}-$2$ is a versatile tool to identify \hi sources, its detection efficiency depends on several parameter settings. For example, the size of the smoothing kernels or the linking length can substantially influence the detection volume in a spectral cube. Moreover, the optimal settings of these parameters could be different for different types of galaxies. For example, a larger smoothing kernel size will have less sensitivity in detecting small galaxies compared to larger galaxies. Further, the flux threshold (the threshold for labeling valid pixels after smoothing) also is a critical parameter, which might have different optimized values for dwarf and spiral galaxies. A higher threshold might recover galaxy properties more accurately in brighter galaxies, whereas the higher threshold might not detect a faint galaxy altogether. Because all types of galaxies exist within a survey volume, different sets of optimized parameters should be used to efficiently recover different kinds of galaxies. 

We use the $40$ GB data cube and the associated truth catalog to identify the best {\sc SoFiA}-$2$ input parameters for detecting sources in the SKA \hi spectral cube. 

The sizes of the smoothing kernels (in all three dimensions) are one of the most important key parameters influencing the detection of faint signals coming from distant galaxies. A kernel much larger than the galaxy size would include more noise than signal while smoothing, reducing the overall SNR. On the other hand, a smaller kernel than the galaxy size would include less signal limiting the maximum attainable SNR. Ideally, an optimum kernel size would match the galaxy size (angular), providing maximum SNR. This requires many optimum kernels, almost as many galaxies in a spectral cube (as the angular sizes/spectral extents would be different). Because we do not have priory information on the galaxy position in a spectral cube, one needs to search the whole cube with all sorts of kernels. This is computationally highly expensive. Instead, we try to identify a set of kernels that reasonably work across different galaxies present in the cube. 

To do the same, we run {\sc SoFiA}-$2$ on the development data cube with different kernel sizes and estimate the detection rates for each kernel. A detection is when the coordinates of a galaxy recovered by {\sc SoFiA}-$2$ matches that of the truth catalog within a beamwidth. The detection rate is determined by the ratio of the number of detections to the total number of galaxies in the truth catalog. This detection rate reflects how efficiently a kernel can detect galaxies in a spectral cube.  

As there is no asymmetry in the spatial direction, \textit{i.e.}, the angular galaxy size in RA and DEC direction have no bias due to random position angle, we chose our spatial kernels to be symmetric. The kernel size in the frequency direction is determined independently. We chose spatial kernel sizes in the range between 3 to 9 pixels, which is equivalent to 8.3$^{\prime \prime}$ to 25.2$^{\prime \prime}$ angular size on the sky. We chose kernel sizes between 3 to 15 channels in the frequency direction, equivalent to $\sim 18 - 90$ km/s. Our choices are motivated by the range of physical extent (both in spatial and frequency direction) of the galaxies. 

In Figure~\ref{fig:eff}, we plot thus obtained detection rates for our chosen kernel ranges. The detection rate and the best-optimized kernel would depend on the brightness of the galaxies. Hence, we divided our whole sample into three flux bins to capture its variation. We evaluate the detection rates for every flux bin (as quoted at the top of every panel in the top row), represented by different horizontal panels in the figure. The color bars indicate detection efficiencies. As can be seen, a spatial kernel size between 3 to 5 and a frequency kernel size between 7 to 15 works reasonably well for all the flux bins.

\begin{figure*}
     \centering
     \includegraphics[scale=0.21]{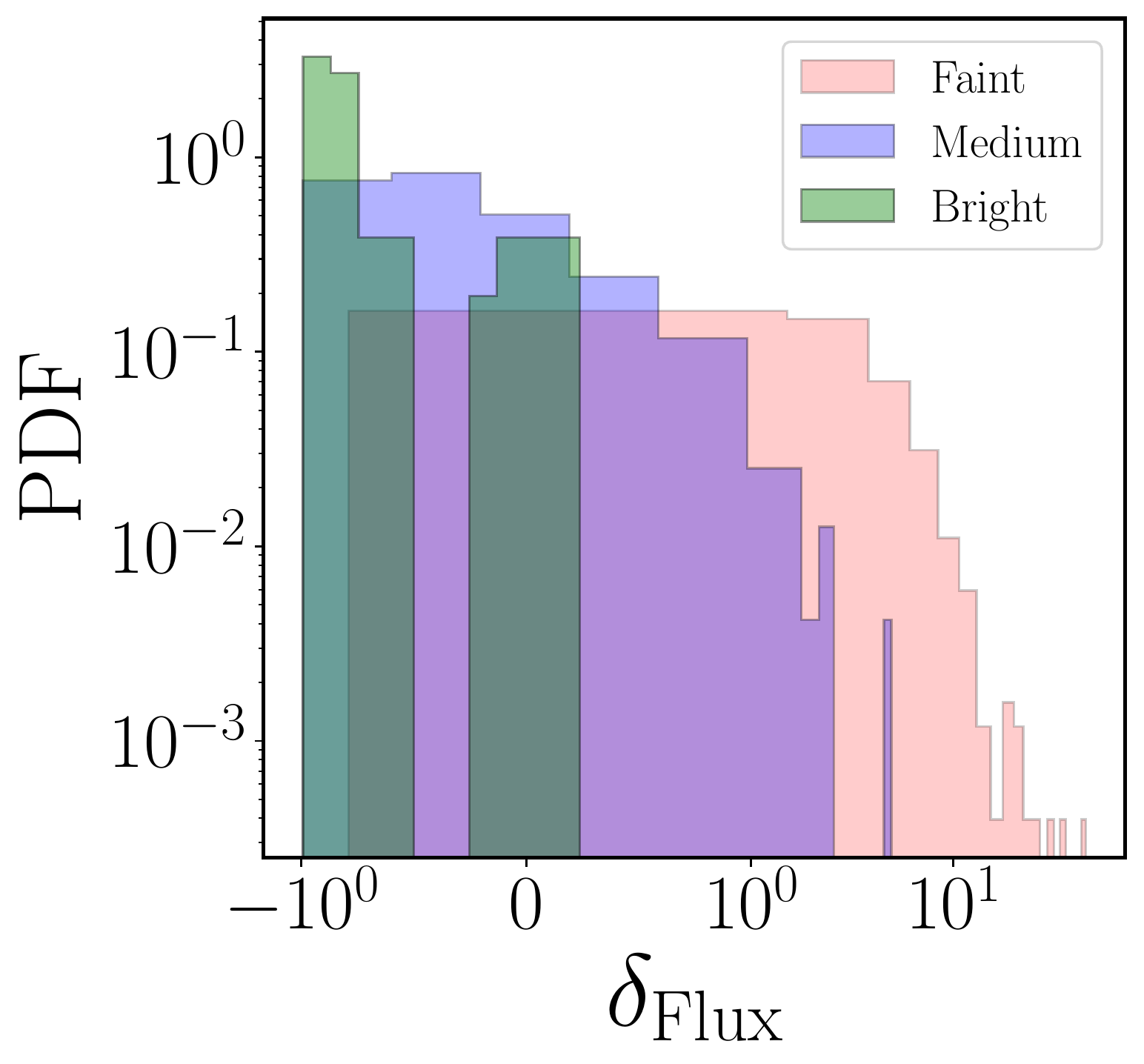}
     \includegraphics[scale=0.21]{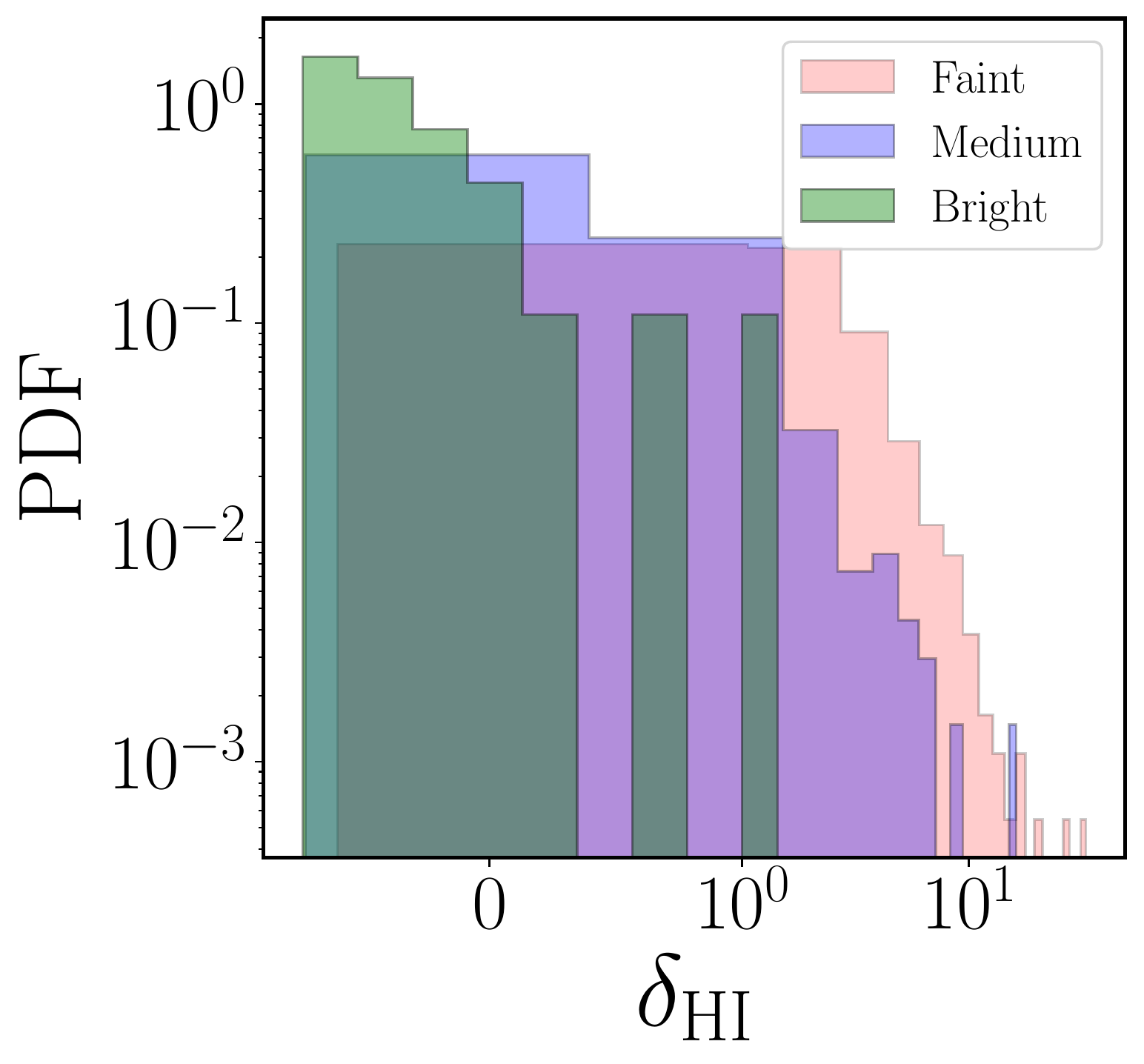}
     \includegraphics[scale=0.21]{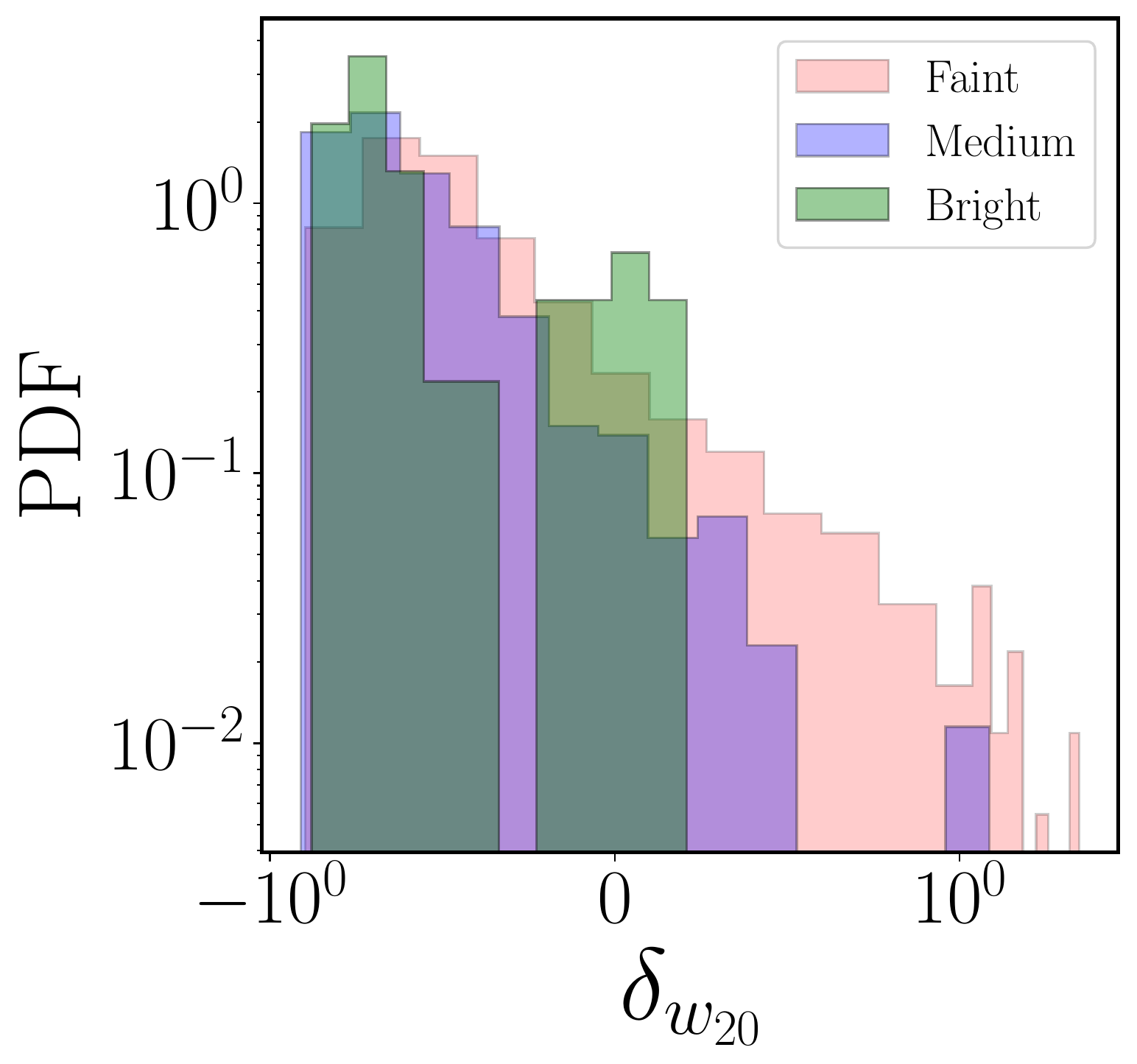}
     \includegraphics[scale=0.21]{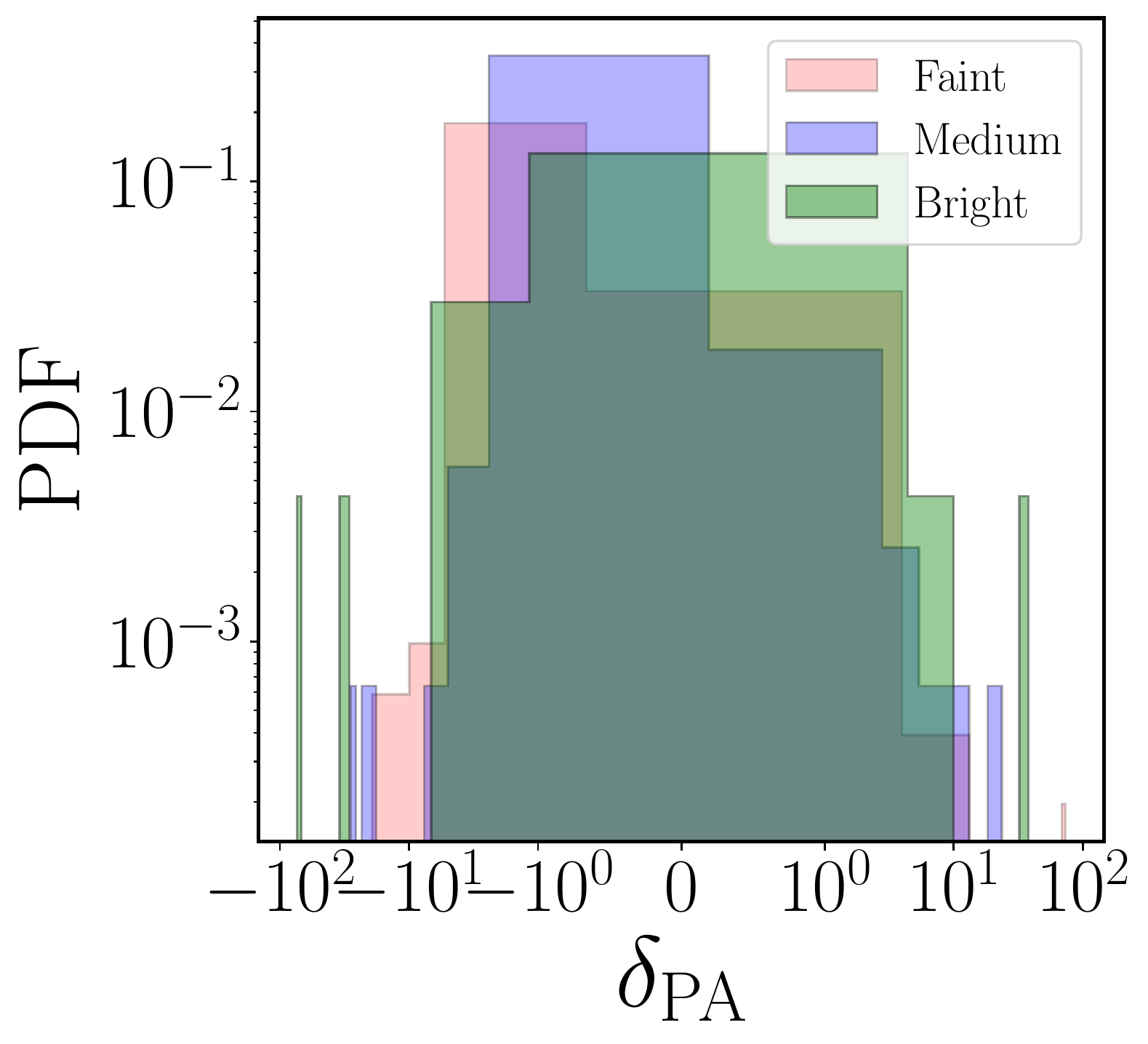}
     \includegraphics[scale=0.21]{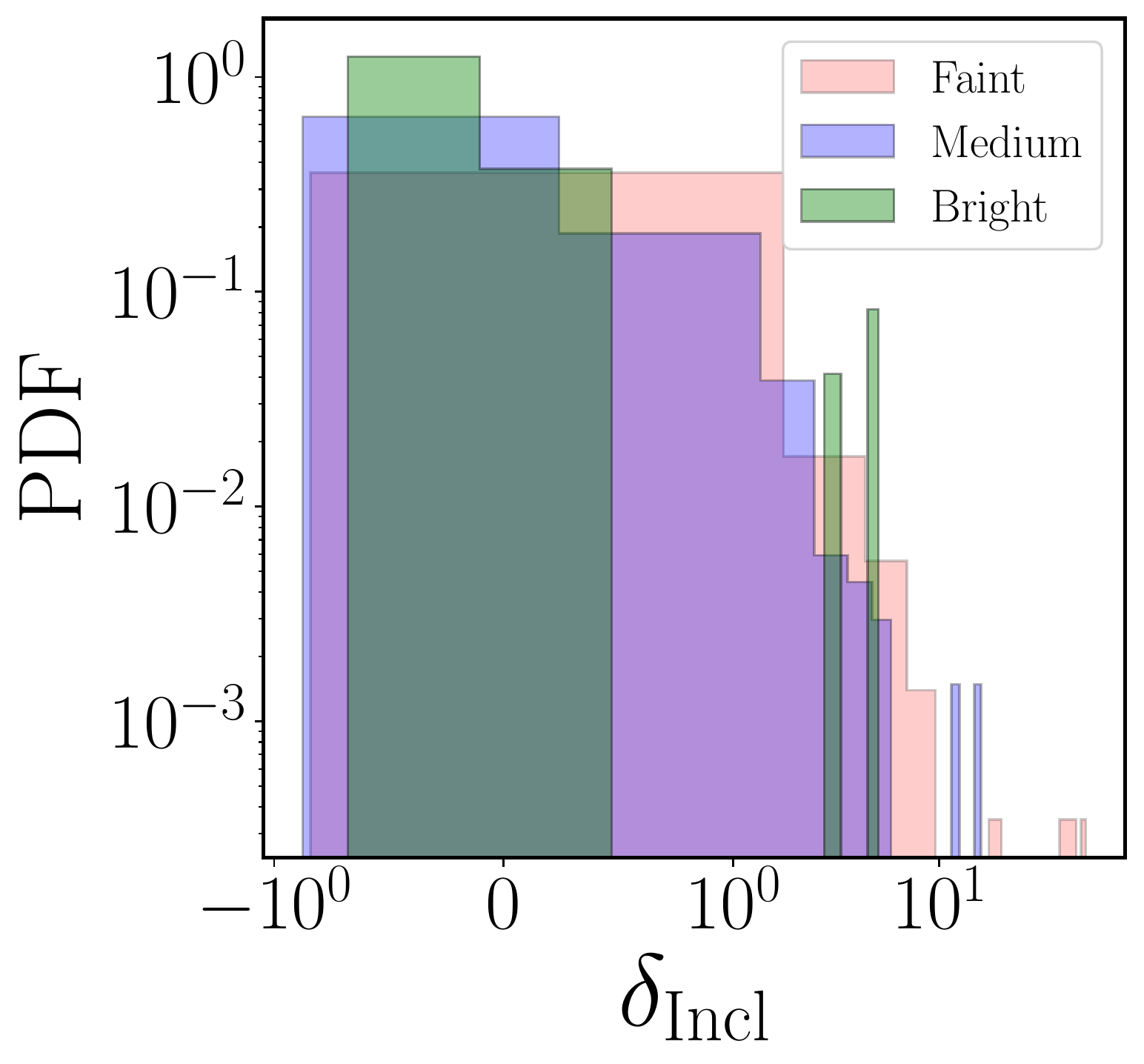} \\
     \includegraphics[scale=0.21]{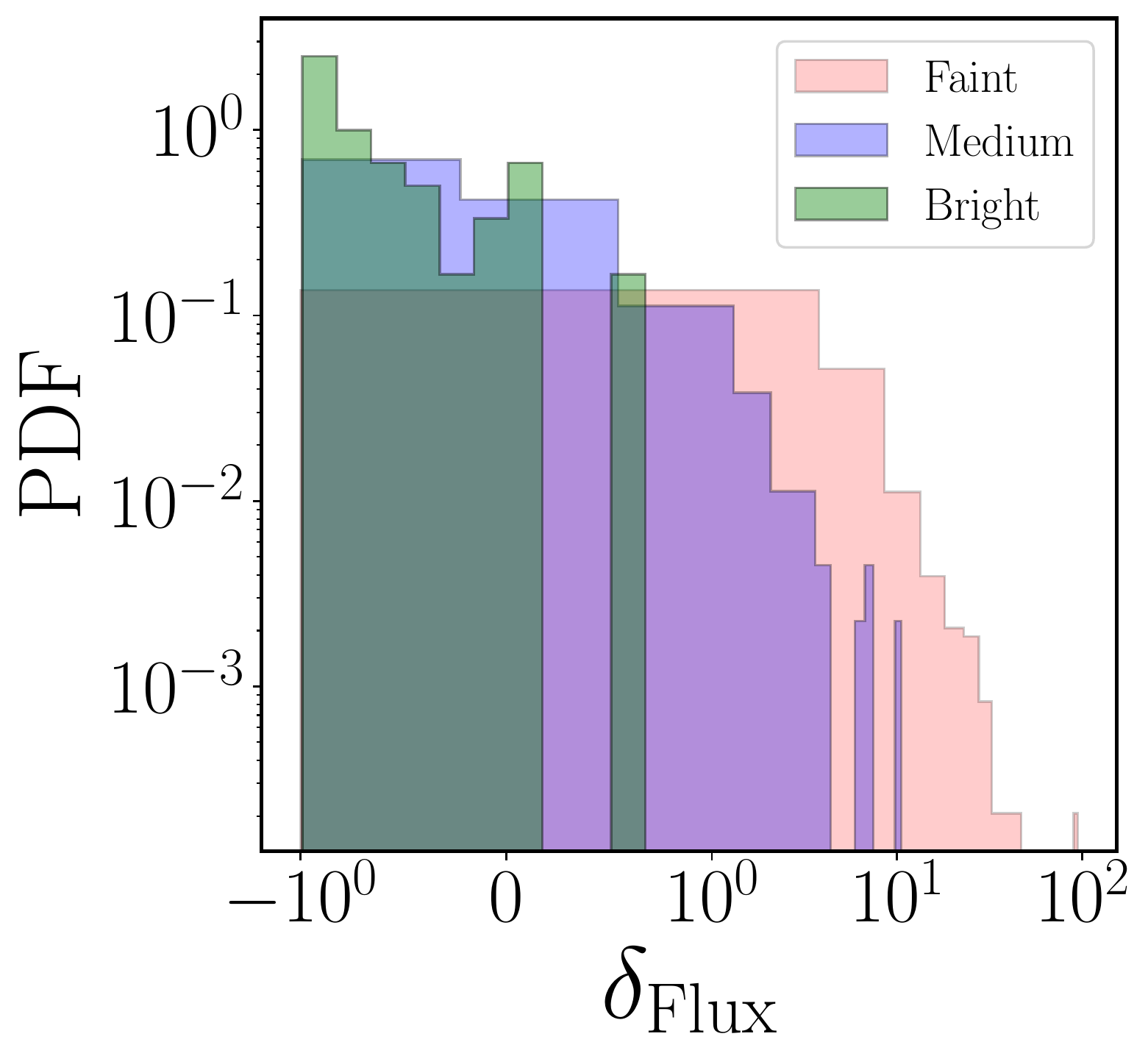}
     \includegraphics[scale=0.21]{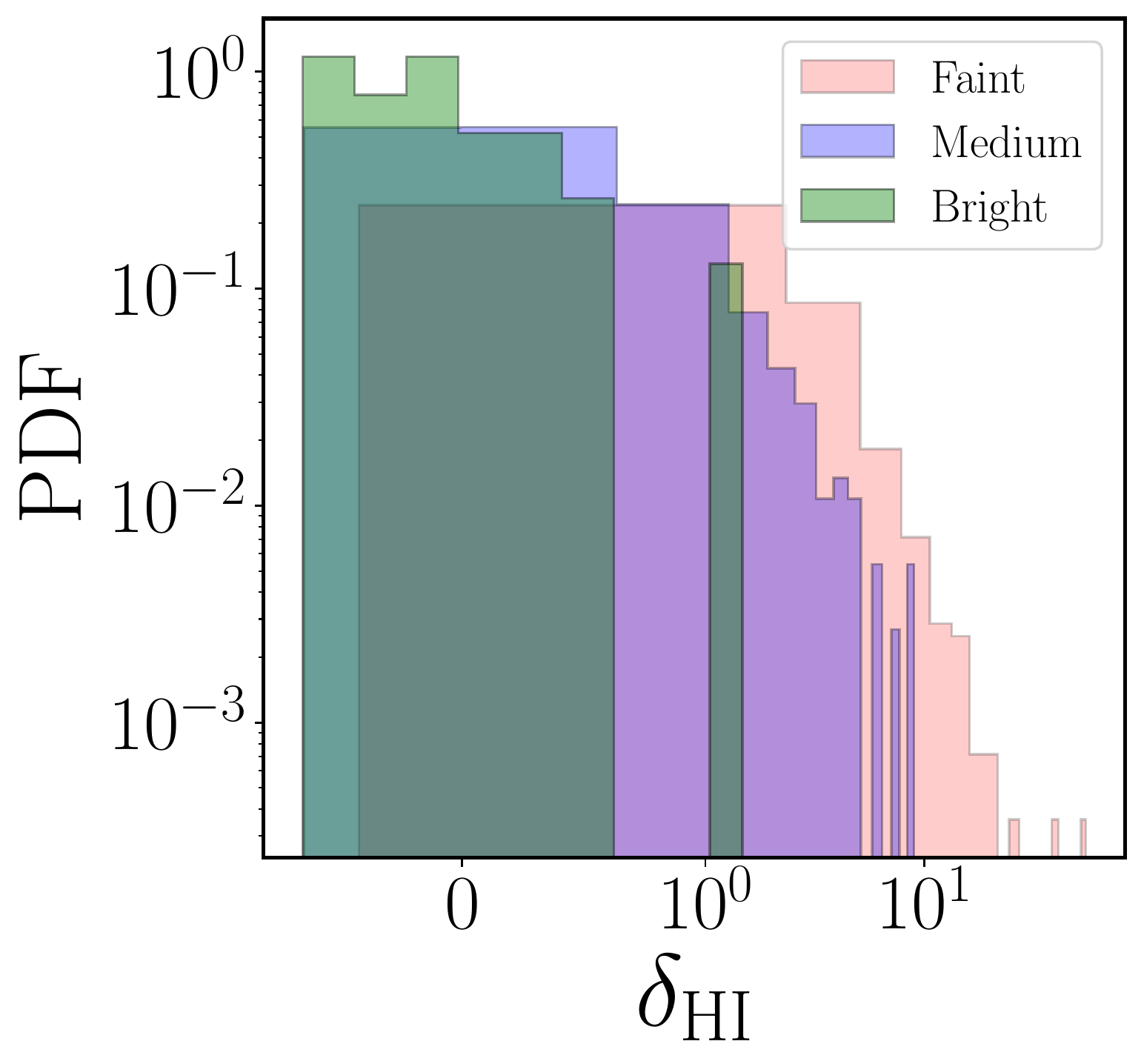}
     \includegraphics[scale=0.21]{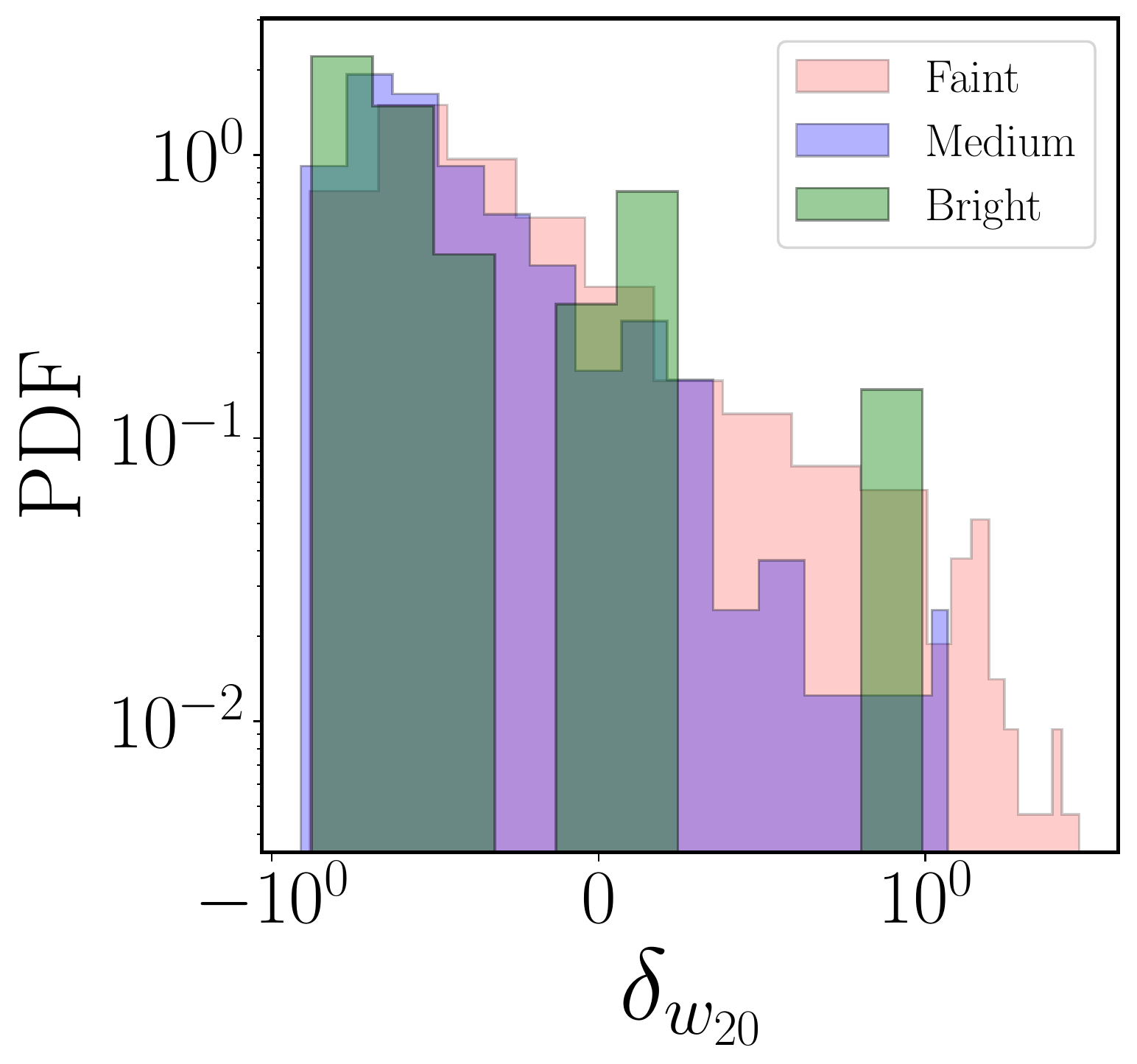}
     \includegraphics[scale=0.21]{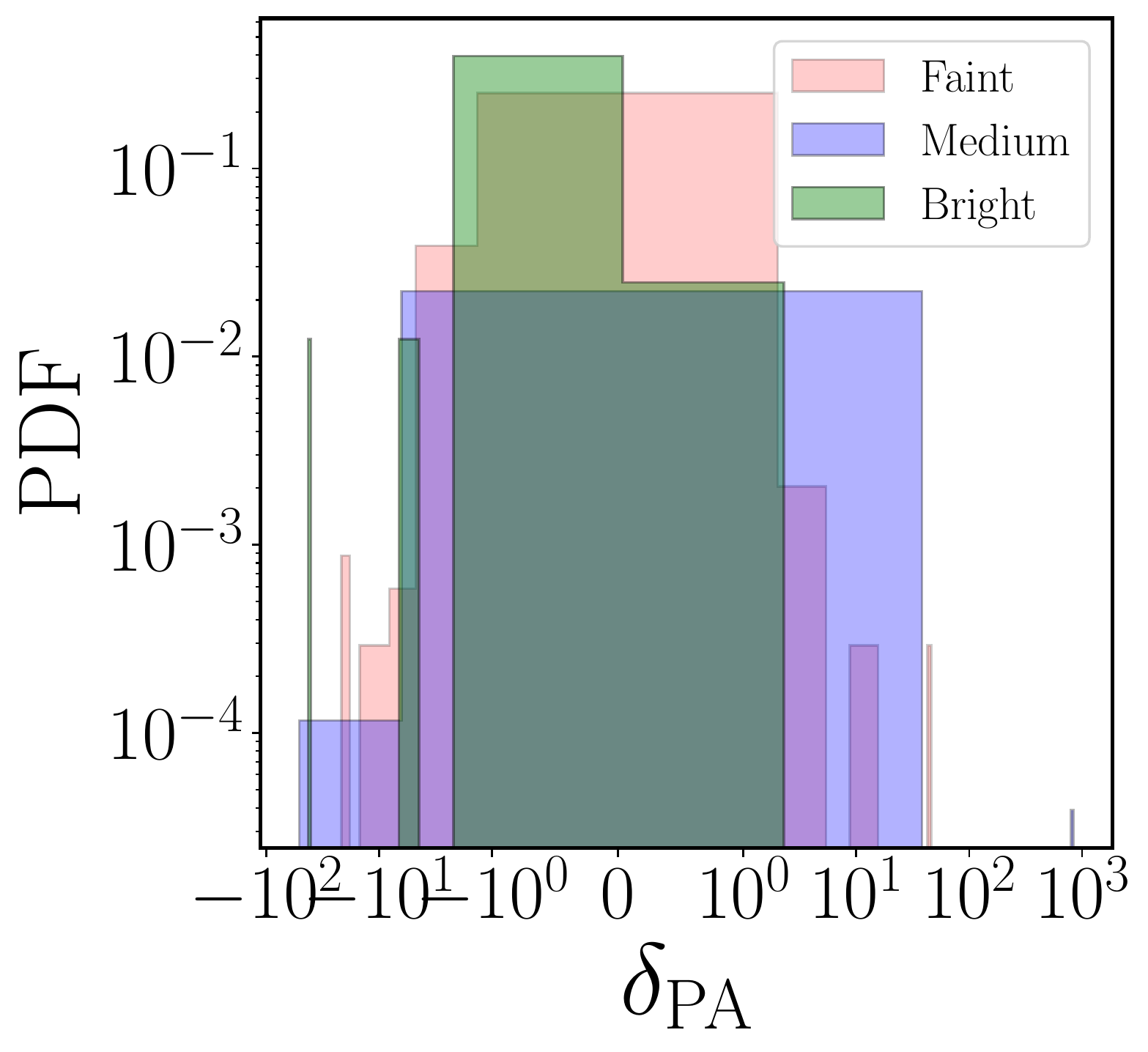}
     \includegraphics[scale=0.21]{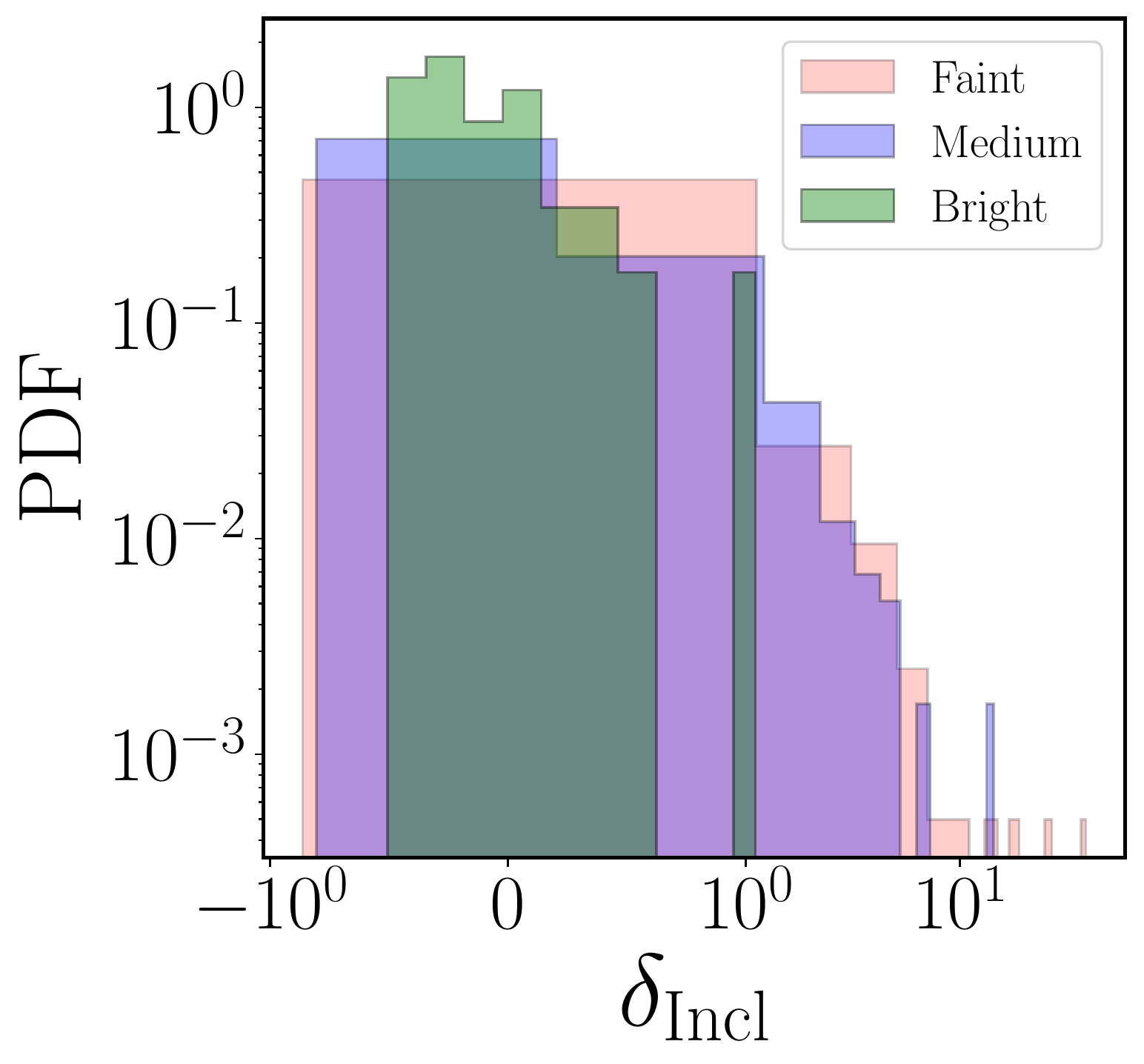} \\
     \includegraphics[scale=0.21]{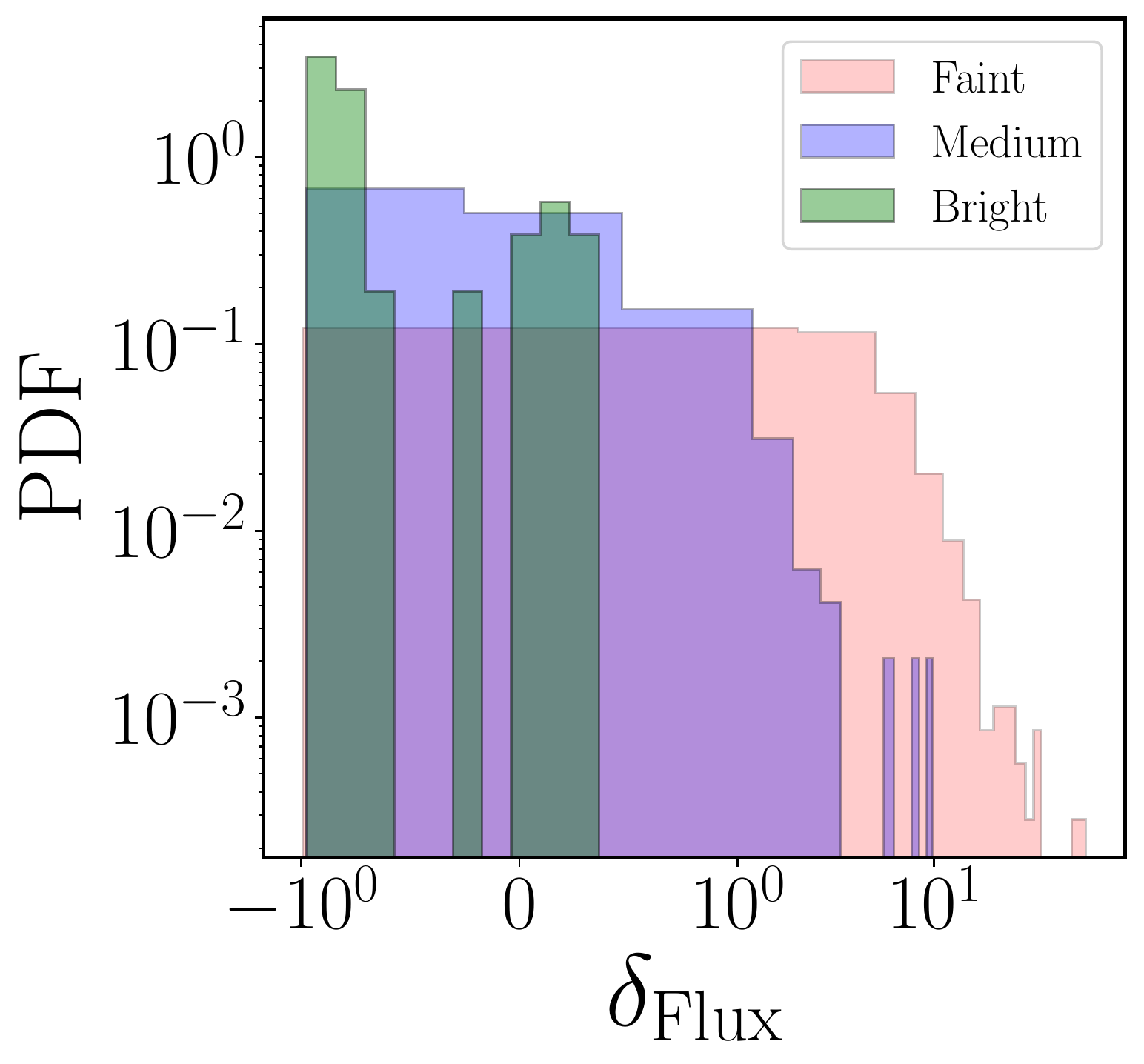}
     \includegraphics[scale=0.21]{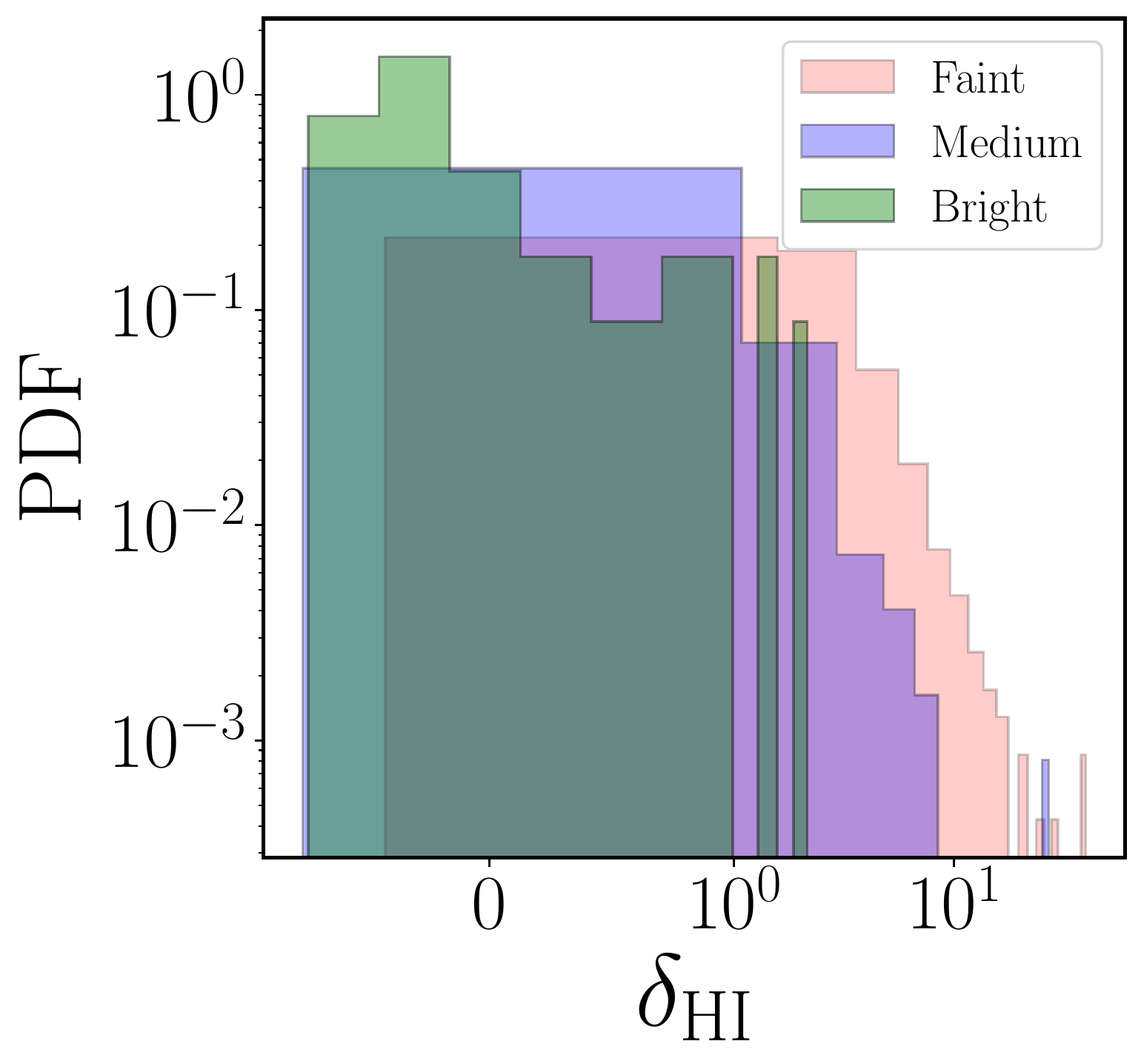}
     \includegraphics[scale=0.21]{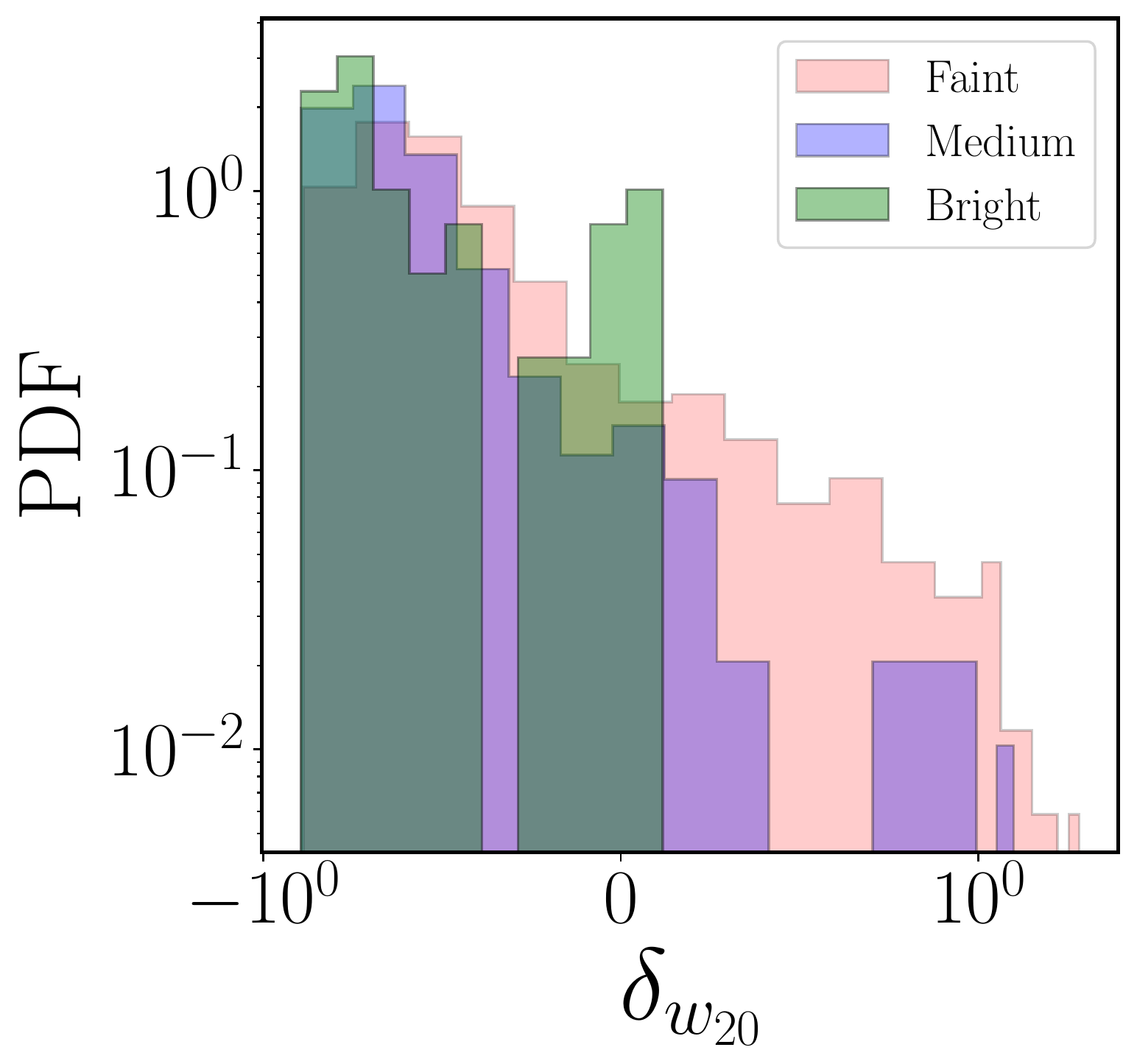}
     \includegraphics[scale=0.21]{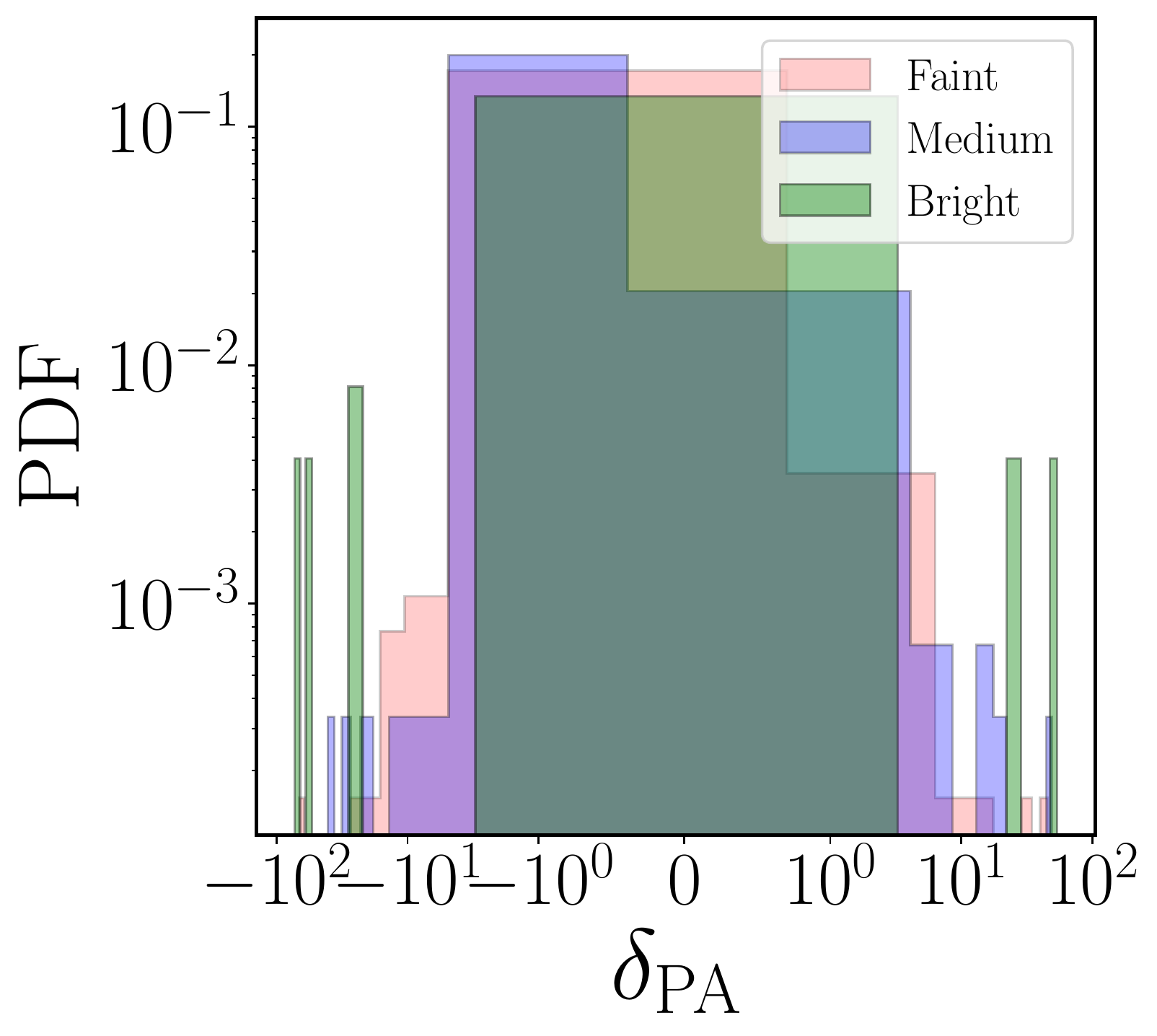}
     \includegraphics[scale=0.21]{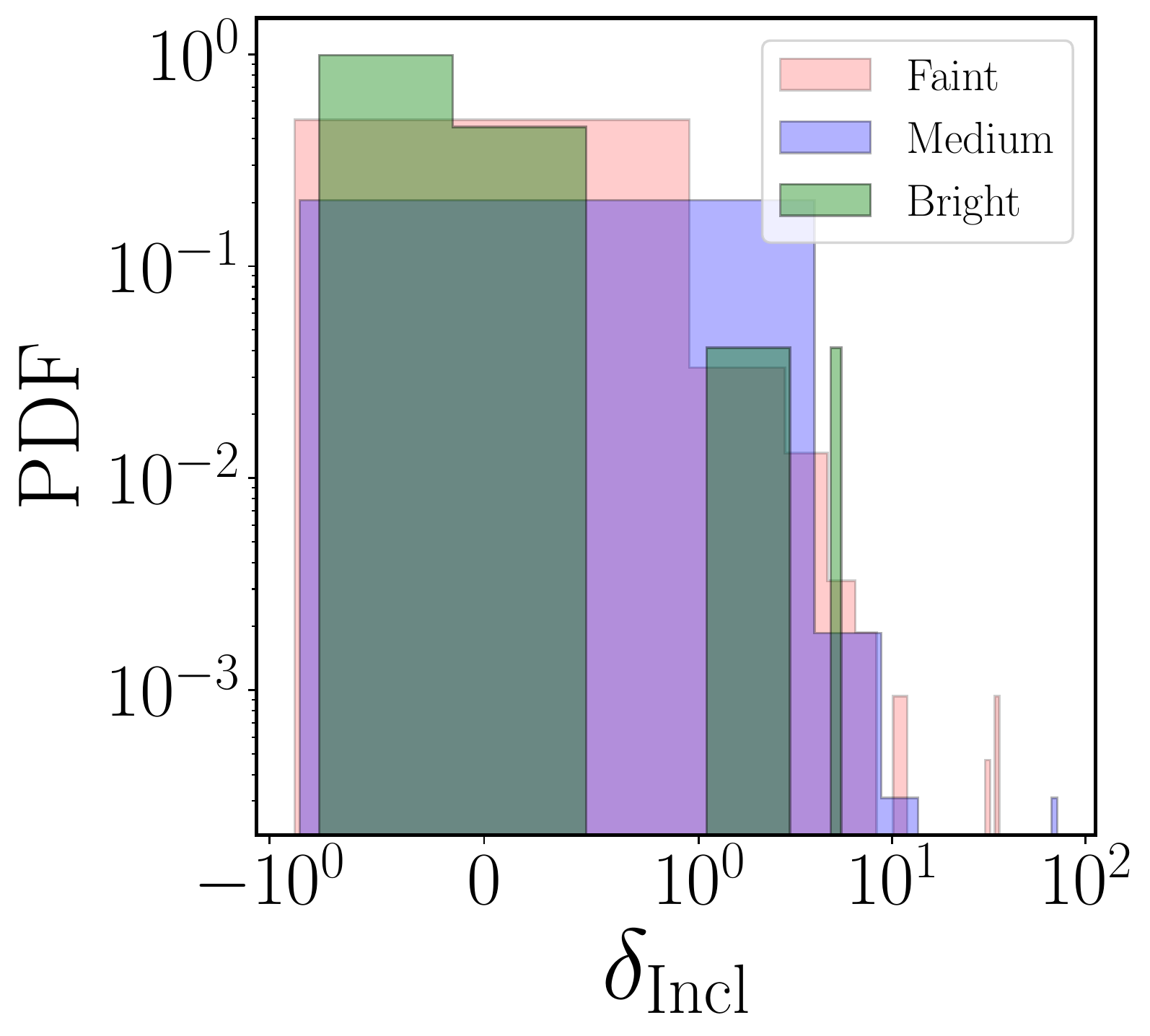}
     
    \caption{Shows the fractional deviations between the recovered and the true values (from the truth catalog) of different galaxy properties. Three colors represent three bins having galaxies with different brightness. Horizontal panels show the PDFs for different properties, whereas vertical panels show the PDFs for different kernels. The top, middle, and bottom panels show the results for spatial kernels, 3,3,5, and frequency kernels, 9,15,9. The deviations shown here are estimated for a 3$\sigma$ threshold.}
    \label{fig:dev_all}
 \end{figure*}
 
  \begin{figure*}
     \centering
     \includegraphics[scale=0.21]{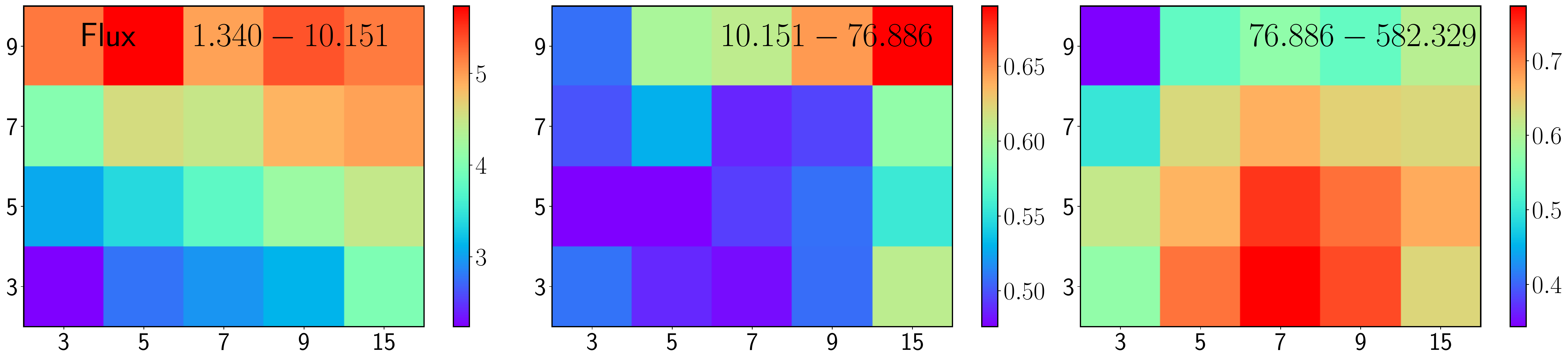}
     \hspace*{-0.55cm}
     \includegraphics[scale=0.21]{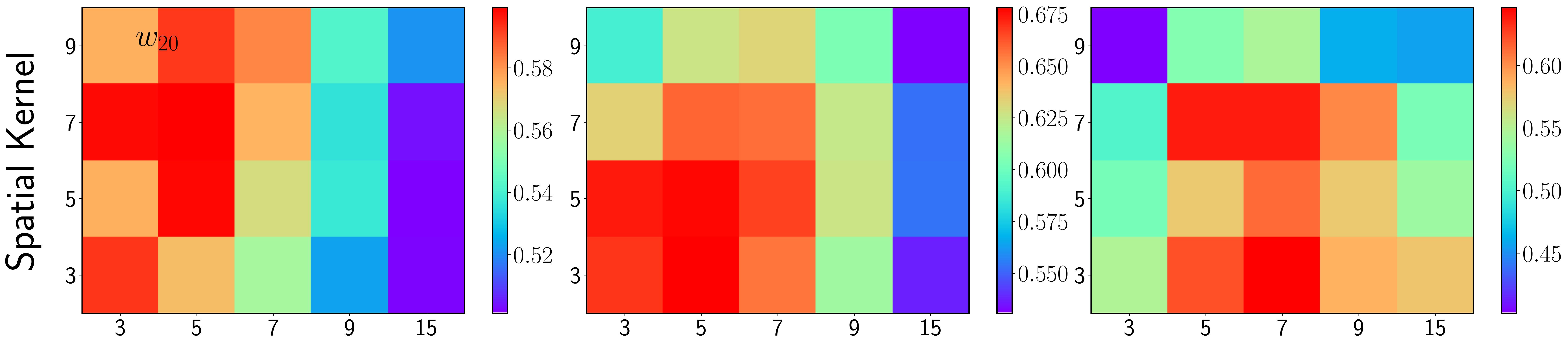}
     \includegraphics[scale=0.21]{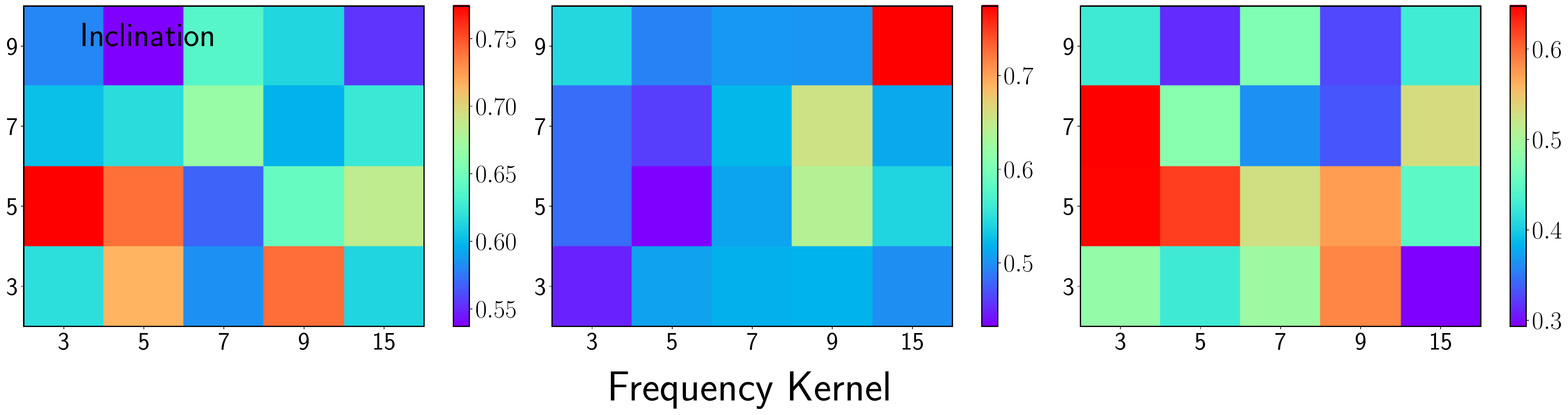}
     \caption{Shows the median fractional deviations of the recovered and the true galaxy properties. Horizontal panels show the deviations for galaxies with different brightness. The top, middle, and bottom panels are for flux, $w_{20}$, and inclination, respectively. See the text for more details.}
     \label{fig:dev_matrix_all}
 \end{figure*}
 
 \begin{figure*}
     \centering
     \includegraphics[scale=0.21]{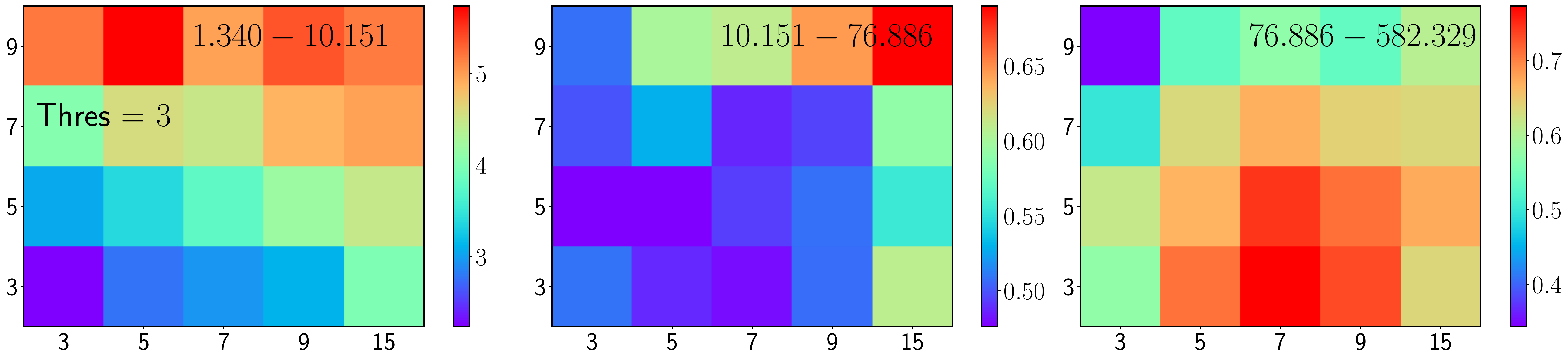}
     \hspace*{-0.55cm}
     \includegraphics[scale=0.21]{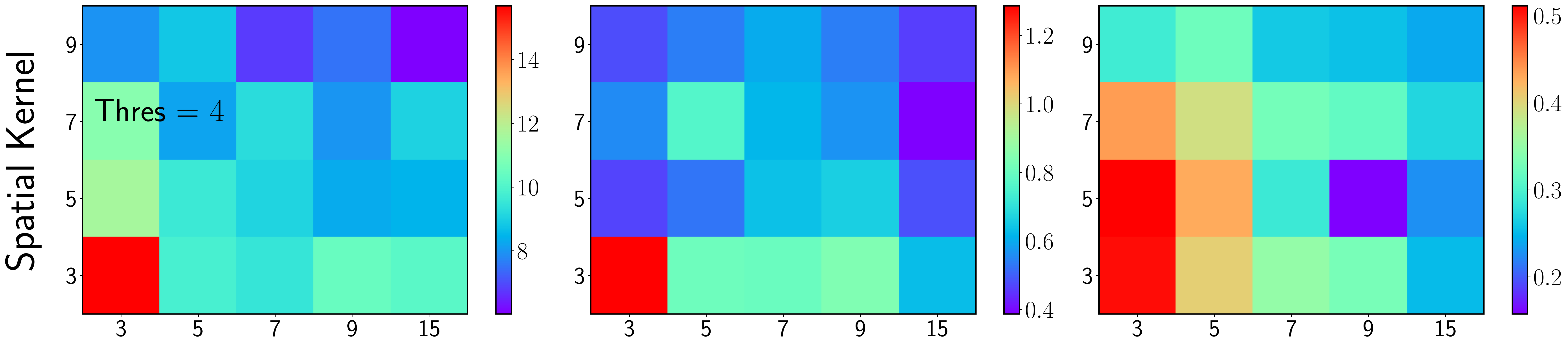}
     \includegraphics[scale=0.21]{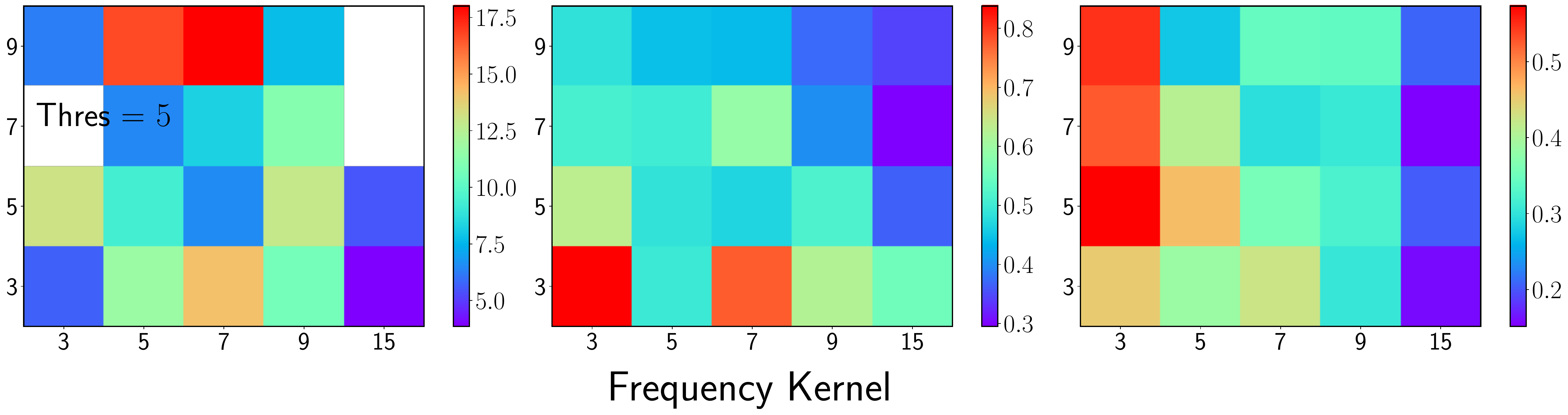}
     \caption{Shows the fractional median deviation of the recovered flux as a function of the chosen threshold. The horizontal panels show the results for galaxies with different brightness. The top, middle, and bottom panels are for thresholds 3,4 and 5, respectively. See the text for more details.}
     \label{fig:dev_matrix_all_threshold}
 \end{figure*}

This exercise was performed using a threshold of 3. As discussed earlier, this threshold decides if a pixel should be identified as a valid one or not after smoothing \citep[see,][for more details]{Serra_2012}. Increasing this threshold would increase the accuracy in recovering galaxy properties at the cost of detection rate. To compare how detection rates can change with increasing thresholds, we perform the same exercise with threshold values 4 and 5. The vertical panels of Figure~\ref{fig:eff} show the results. As we can see, with an increased threshold, the detection efficiencies decrease significantly. This happens because low flux faint galaxies get excluded at a higher threshold. Because of the larger number of faint galaxies (i.e., more number of dwarf galaxies) in an unbiased survey volume, the efficiency dramatically decreases with a slight increase in the threshold. 

Further, we notice that unlike threshold 3, no reasonably good kernels can be found for higher thresholds, which works for all flux ranges. Given these results, we chose to use a spatial kernel of size 3 pixels and a frequency kernel of size 9. We also chose to work with a threshold of 3. With these {\sc SoFiA}-$2$ settings, we run our routine on the data cube to build the final catalog. Here we stress that as the truth catalog for the developmental data set is available, we use it to characterize our routine. We will use our routine with these optimized settings for an observed data cube to detect \hi sources. We note that we recover only 15-25\% of galaxies listed in the truth catalog with these settings. Though it might seem to be a low efficiency, we emphasize that this is relative. Not all the galaxies in the truth catalog need to be detected in the spectral cube. There could be galaxies so faint that they would be buried under noise, and no detection technique would be able to recover them. In that sense, a detection rate of 15-25\% could be reasonable \cite{skadc2}. In this context, we would like to mention that our results are consistent with those obtained by other SKA-DC2 teams.

For cross-matching our detections (\textit{i.e.}, counting a valid detection by our routine), we have only used position, \textit{i.e.}, RA, DEC, and frequency. However, the other recovered physical parameters, e.g., total flux, galaxy size, inclination, etc., are also critical for estimating galaxy properties. We note that identifying a galaxy by an optimized kernel does not guarantee the optimized recovery of these parameters by the same kernel. Further, different kernels might best suit to recover different properties for the same set of galaxies. A small kernel might better identify the center of a galaxy, but a larger kernel might work better to recover the total flux. Ideally, optimized kernels should be identified for every galaxy parameter (starting from size, flux to inclination). An observed cube then should be passed through all these optimized kernels. However, this demands significantly higher computation, as the data cube has to be searched for many times for all the different kernels. Here, we only use a single kernel (\textit{i.e.}, 3 and 9) to detect sources with reasonable confidence. Nonetheless, we characterize the accuracy of our routine in recovering the galaxy properties. 

\begin{figure}[ht!]
    \centering
    \includegraphics[width=0.45\textwidth]{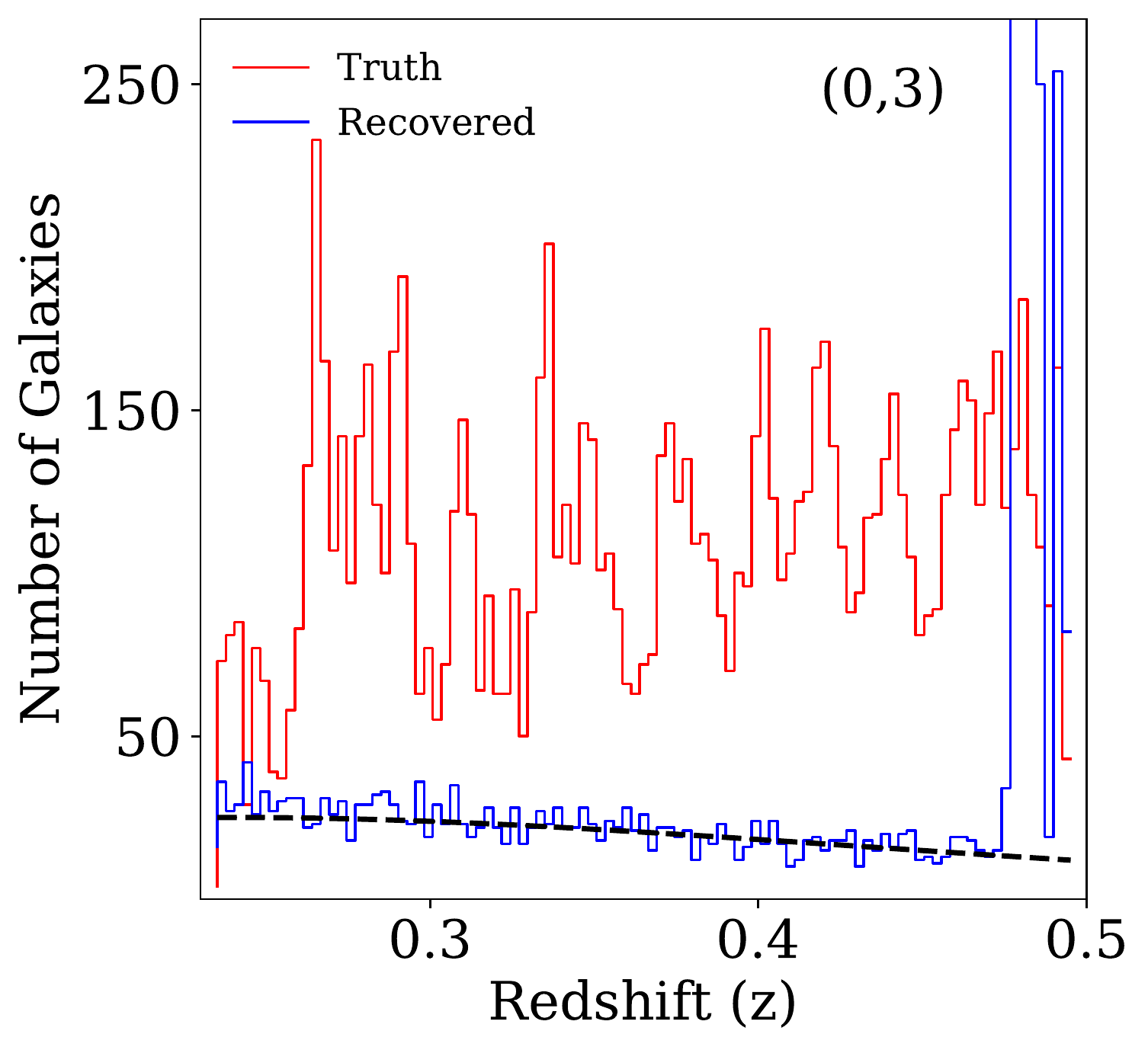}
\caption{Shows the number of detected galaxies as a function of redshift using our detection method. The red histogram shows the number of galaxies in the truth catalog, whereas the blue histogram represents the number of detected galaxies. The detection rate is estimated for a spatial kernel 0 and a frequency kernel 3. The survey's selection function is shown by the black dashed line. As can be seen from the figure, our overall number of detection (by the kernel (0,3)) matches well with the survey selection function. See the text for more details.}
    \label{fig:nz}
\end{figure}

In Figure~\ref{fig:dev_all} we plot the Probability Distribution Functions (PDFs) of the deviations of the recovered physical properties of the detected galaxies from the truth values. Horizontal panels refer to different properties. As can be seen from the figure, the fractional deviations in our estimated properties are large, and sometimes it could be a factor of a few. These deviations could originate due to two reasons. Firstly, some of the signals could be lost intrinsically due to noise. This is fundamental and cannot be solved by changing detection techniques, which is expected to affect the total integrated flux, \HI~mass, and the galaxy size more than other parameters. Secondly, due to poor choice of kernel, or thirdly both. As further can be seen from the figure, the deviations are more for fainter galaxies (orange histograms) than the brighter ones. This is expected since a significant fraction of their flux/mass is in the diffused state with low SNR, easily buried under the noise floor. Due to computational limitations, in general, we chose to use a single kernel to operate on the whole cube containing different types of galaxies. However, the accuracy in the recovered galaxy properties would depend on the choice of kernel sizes. We performed the above exercise with two other kernels, \textit{i.e.}, 3,15 and 5,9, to test how the PDFs of the deviations change. We show the results in the vertical panels of the figure. As can be seen, with different kernels, the PDFs change considerably. This indicates that the choice of a proper kernel while estimating the galaxy properties is critical. 

To investigate this in more detail, we ran our source finding routine over the entire cube with a variety of kernels to search for the best one to recover the galaxy properties more accurately. For these kernels, the deviations would be minimum. In Figure~\ref{fig:dev_matrix_all} we plot these deviations. The three horizontal panels are for galaxies with different total fluxes (faint, moderate, and bright). The top, middle, and bottom panels indicate the deviations for flux, $w_{20}$, and inclination. As can be seen, the deviations are very different for different kernels. For example, a smaller kernel seems better suited for recovering total flux in faint galaxies (top panels). However, the same kernel does not seem to work best to recover $w_{20}$ (middle panels). This further strengthens the idea that optimal kernels are specific to galaxy properties and the best kernels need to be searched for each of them. Thus, a thorough search on the kernel space should be done while recovering galaxy properties. 

The recovered galaxy properties could be sensitive to the inclination for the exercise mentioned above. Hence, considering one individual galaxy in each horizontal panel might bias our result (as the inclinations might be different). Thus, in Figure~\ref{fig:dev_matrix_all}, we calculate the median of the deviations for all the detected galaxies within a flux bin. The considerable number of detected galaxies within a bin is expected to have the full range of inclinations, removing the bias. 

Further, this exercise is done with a {\sc SoFiA}-$2$ threshold of $3$. Though this is a widely used standard cut-off, it can substantially include noise as a signal. This will corrupt the faint galaxies significantly where the peak flux could be just a little higher than $3\sigma$. Deviations would be large for these galaxies, and no kernel would be best to recover the correct properties. To check the same, we perform the same exercise mentioned above with thresholds of $4$ and $5$. The results are plotted in Figure~\ref{fig:dev_matrix_all_threshold}. The horizontal panels refer to faint, moderate, and bright galaxies; the vertical columns indicate the deviation for thresholds of $3$, $4$, and $5$ (top, middle, and bottom). As can be seen from the figure, with increasing threshold, the deviations reduce. For example, for the brightest galaxies (right panels), the fractional deviation decreases from $\sim 40\%$ to $\sim 20\%$ as the threshold increases from $3$ to $5$. The same trend can be seen for the moderately bright galaxies (middle panels) as well. However, for faint galaxies, this trend is disrupted. As we discussed, a higher threshold might exclude a large amount of flux in these galaxies leading to large deviations. This, in turn, can suppress the trend seen for the relatively brighter galaxies.

Next, to investigate the overall detection efficiency of our routine, we calculate the detection rates as a function of redshift. In Figure~\ref{fig:nz} we plot the number of detected galaxies as a function of redshift. As can be seen from the figure, the input number of galaxies in the cube is roughly constant (red histogram) as a function of redshift. This is because the (possible) evolution of the \hi~mass function is not implemented while simulating the spectral cube. However, the detection rate is expected to decline with redshift due to the change in the detectability of minimum \hi~mass. This can be quantified using a survey selection function, \textit{i.e.}, the number of expected detections given the survey depth and coverage. For our data set, the minimum detectable \hi~masses at the nearest ($z=0.235$) and furthest ($z=0.495$) redshifts are $\sim 1.3 \times 10^{9} ~ {\rm M}_{\odot}$ and $\sim 1.7 \times 10^{9} ~ {\rm M}_{\odot}$ respectively. The coverage of the simulated survey is $1$ square degree. With these survey depth and coverage, we estimate the survey selection function and found that it matches our detection rates reasonably well for the kernel $(0,3)$. However, we note that the detection rates vary significantly for different kernels. We emphasize that the calculation of our selection function is purely based on the sensitivity of the survey and its coverage. We do not consider the effect of possible imperfections in the data, e.g., RFI, imperfect subtraction of continuum, etc. Further, we note that our routine detects an unusually large number of sources at the highest redshift bins (lowest frequency bins). This is probably due to the inclusion of simulated RFIs in the data at the lowest frequencies \citep[see,][for more details]{skadc2}.

\begin{figure*}
    \centering
    \includegraphics[width=0.9\textwidth]{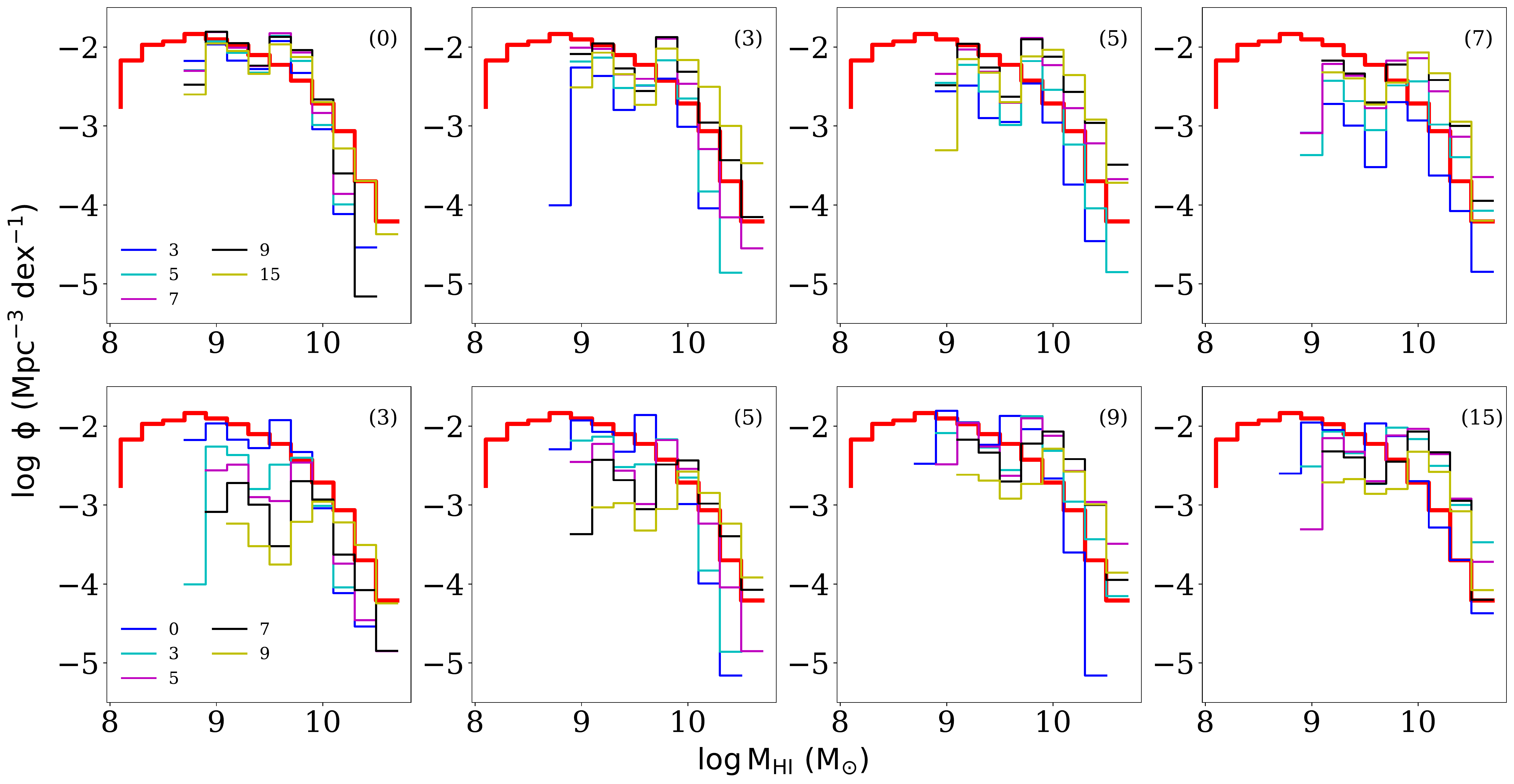}
    \caption{Shows the derived \hi~mass function for different adopted spatial and frequency kernels. In each panel, the red histogram represents $\phi \thinspace ({\rm M}_{\rm HI})$ derived from the truth catalog. The top panels show the variation of $\phi \thinspace ({\rm M}_{\rm HI})$ with different spatial kernels (as quoted on the top right in each panel). Each individual panel shows the same for different frequency kernels as quoted in the legend. The bottom panels show the variation of $\phi \thinspace ({\rm M}_{\rm HI})$ with different frequency kernels. As can be seen, $\phi \thinspace ({\rm M}_{\rm HI})$ varies significantly with the choice of kernels, with the kernel $(0,3)$ providing a reasonable match to the truth $\phi \thinspace ({\rm M}_{\rm HI})$ (top left panel). See the text for more details.}
    \label{fig:himf_spatial}
\end{figure*}

To further investigate the detection rates at different mass bins, we use our source catalogs to estimate the \hi~mass function $\phi \thinspace ({\rm M}_{\rm HI})$ which is shown in Figure~\ref{fig:himf_spatial}. We found that the recovered mass function significantly depends on the choice of kernels. Smaller kernels tend to detect low-mass galaxies well, whereas larger kernels detect larger galaxies better. We also fund that $\phi \thinspace ({\rm M}_{\rm HI})$ obtained using kernel $(0,3)$ matches the truth one better than other kernels (top left panel). However, for this kernel also, the $\phi \thinspace ({\rm M}_{\rm HI})$ shows considerable deviation from the truth value. Hence, a combination of different kernels suited for different types of galaxies would be ideal for galaxy detection and parameter estimation in an \hi~spectral cube. 

As can be seen from these exercises, the choice of kernels is crucial both in detecting and recovering galaxies in an \hi spectral cube, especially for fainter galaxies. Ideally, individual galaxy candidates (after the detection) should be passed through a host of kernels to identify property-specific optimized kernels. Thus a rigorous study is required to create an ideal kernel template for particular sets of galaxies, e.g., with specific flux and inclination range, etc. For our study, we only use a single kernel for all purposes. However, we emphasize that as preliminary work, our routine is mainly focused on detecting potential galaxies in the \hi spectral cube. We plan to develop a routine that will inspect these individual detections rigorously and recover the properties with the optimum kernels/{\sc SoFiA} settings from a predetermined template matrix.


\section{Summary and Future scope}
\label{summary} 

Hydrogen, being the most abundant baryonic component in the Universe, is the most promising tracer to study galaxy formation and evolution. Its hyperfine transition line of $1420~{\rm MHz}$ is an excellent tracer to map \hi content and its dynamics within the galaxies at $z\sim1$. Thus, several surveys such as HIPASS, ALFALFA, and CHILES have conducted observations to provide constraints over \hi  mass function from the galaxies. However, they were limited to the local Universe ($z<0.4$) due to limited sensitivity. SKA and its precursor facilities like the MeerKAT and ASKAP are expected to have enhanced sensitivity to push the threshold of \hi detection from galaxies to higher redshifts. The precursors are currently carrying out several deep surveys such as MHONGOOSE, MIGHTEE, LADUMA, WALLABY, FLASH, DINGO, etc. to observe galactic \hi contents up to redshift $z\lesssim 1$. SKA will add to the findings of these surveys by further observing the \hi within the galaxies and the infalling gas with much higher resolution. Thus, it will be able to constrain the HIMF and Tully-Fisher relations of the galaxies with higher fidelity up to $z\approx 1$. However, the data volume for surveys with SKA will be $\sim$few TBs even after some real-time processing for data volume reduction. Analyzing such large data volumes is challenging with the currently available tools and computational resources. Hence, there is need in the community for developing more sophisticated and automated data analysis pipelines to handle the expected large data volumes. This paper discusses our efforts to analyze such big data using traditional and modern methodologies.

We have developed our algorithm to analyse large ($\sim {\rm TB}$) spectral radio-data cubes. We have tested our methods on a $40$ GB (comoving volume $645833.2\,{\rm Mpc}^3$) data cube which was made publicly available by the SKA-DC2. Our results in this paper also is based on the analysis of this $40$ GB data cube. Note that we have already applied our algorithm to the original $1$ TB\footnote{The 1 TB cube was restricted only to the participants through some private computational facilities.} data cube of the SDC2 challenge, in which we had taken part as Team Spardha. The cube was simulated between a frequency range $950-1150~{\rm MHz}$ corresponding to the redshift range $0.235-0.495$, with $6668$ channels along with the frequency axis. Both cubes were simulated between a frequency range $950-1150~{\rm MHz}$ corresponding to the redshift range $0.235-0.495$, with $6668$ channels along with the frequency axis. However, $40$ GB and $1$ TB data cubes span $1~{\rm deg}^2$ and $20~{\rm deg}^2$ respectively on the sky. The \hi signal within the data cubes was simulated to match the \hi mass-functions extrapolated from the observations in the nearby Universe. Also, the data cube was infused with system noise corresponding to $2000$ hours of SKA observations, residual continuum, RFI, systematics, etc., to make the data cube realistic.

We first applied our MPI-based {\sc Python} pipeline which divides the large data cubes into multiple small cubelets and then find galaxies within each cubelet in parallel. The pipeline uses publicly available {\sc SoFiA}-$2$ as its core ingredient to find the \hi sources in the cubelets. {\sc SoFiA}-$2$ works on a traditional smoothing and clipping ({\bf S+C}) algorithm for finding galaxies from a noisy data. Therefore, the smoothing kernel size and the clipping thresholds are the parameters which affects the performance of {\sc SoFiA}-$2$. Note that our pipeline is designed in a way that takes care of the sources which lie in the boundaries between the two cubelets. We first explore how the performance of our pipeline depends on the {\sc SoFiA}-$2$ kernel sizes and threshold by varying the spatial kernel size and frequency kernel size (in terms of pixels) as well as the clipping threshold step wise. We compare the position of detected sources (RA, DEC and central frequency) with those in the true list of source we already have. We have binned the \hi sources into three classes, \textit{i.e.} faint, moderate and bright, based on their fluxes and present results for them separately. We note that, for all flux bins, a spatial kernel size between $3$ to $5$ pixels and a frequency kernel size between $7$ to $15$ pixels works reasonably well at a threshold of $3\sigma$. Increasing the threshold level may recover the source properties (e.g. flux, line-width, inclination, \hi size) accurately (Figure \ref{fig:dev_matrix_all_threshold}), however the detection of the sources in each flux bin drops drastically (Figure \ref{fig:eff}). We also note that there is no such common kernel size combination for which detection efficiency is optimum in all of the flux bin. This indicates that multiple sets of kernels are required for detecting galaxies with widely varying fluxes. This is also true with other source properties such as flux, line width ($w_{20}$), inclination etc. We find that a common kernel size is not sufficient to recover all the source properties as well as the \hi mass function $\phi({\rm M}_{\rm HI})$ with a legitimate accuracy (Figures \ref{fig:dev_matrix_all} and \ref{fig:himf_spatial}). Therefore, one needs to run the pipeline with multiple kernel sizes to accurately detect sources with widely varying fluxes. This will be a time consuming process when the data volume is $\sim$ a few TB. We plan to improve the pipeline by employing an automated routine that will inspect these individual detections rigorously and recover the properties with the optimum kernels/{\sc SoFiA} settings from a predetermined template matrix.

The output from {\sc SoFiA} is not completely reliable due to possibility of large false detection, as the data we are dealing here have smaller SNR than any typical low-redshift observation. Therefore, we are also developing a hybrid pipeline where we plan to use convolutional neural networks (CNNs) as a classifier to remove the false detections in {\sc SoFiA} outputs. We plan to make use of the moment-1 maps, obtained from {\rm SoFiA}, to discard the false detections. We are also in the process of developing modern machine learning based algorithms on our own to detect the \hi sources directly from the noisy data cubes. Our present efforts in this direction indicates that CNNs based on U-Net image segmentation algorithm can be effective. Our primary goal here is to recover galaxy extent more accurately besides its location. Additionally, we are also exploring the application of friends-of-friend (FoF) based algorithm which has potential to find out galaxies using thresholding and linking of data pixels. Note that both the CNN-based as well as FoF-based algorithms are still being developed and tested for their robustness, and it is an ongoing work. We, as a team, are continuously working to develop a data analysis pipeline which will be efficient enough to handle large data volumes generated from the future SKA observations.

\section*{Acknowledgment}
Authors would like to acknowledge SKA India Consortium and IUCAA for the support with the computing facilities. We would also like to thank Nirupam Roy for useful discussions. We acknowledge National Supercomputing Mission (NSM) for providing computing resources of `PARAM Shakti' at IIT Kharagpur, which is implemented by C-DAC and supported by the Ministry of Electronics and Information Technology (MeitY) and Department of Science and Technology (DST), Government of India. We also acknowledge Raman Research Institute for providing their HPC facility. RM is supported by the Wenner-Gren Postdoctoral Fellowship. AKS acknowledges support by the Israel Science Foundation (grant no. 255/18). AM acknowledges Indian Institute of Technology Indore for supporting the research through Teaching Assistantship. 

\bibliography{ref}
\end{document}